\documentclass{aa}
\usepackage{array}
\usepackage{graphicx}
\usepackage{float}
\usepackage{txfonts}
\begin{document}

  \title{Kinematics of coronal mass ejections in the
LASCO field of view}


  \author{Anitha Ravishankar
         \inst{1}
         \and
         Grzegorz Micha$\l$ek\inst{1}
         \and
         Seiji Yashiro\inst{2}
}

\institute{Astronomical Observatory of Jagiellonian University, Krakow, Poland
\and
            The Catholic University of America, Washington DC 20064, USA          \\
             \email{anitha@oa.uj.edu.pl}
             }




 \abstract
  {In this paper we present a statistical study of the kinematics of 28894 coronal mass ejections (CMEs) recorded by the \emph{Large Angle and Spectrometric Coronagraph} (LASCO) on board the \emph{Solar and Heliospheric Observatory} (SOHO) spacecraft from 1996 until mid-2017. The initial acceleration phase is characterized by a rapid increase in CME velocity just after eruption in the inner corona. This phase is followed by a non-significant residual acceleration (deceleration) characterized by an almost constant speed of CMEs. We demonstrate that the initial acceleration is in the range 0.24~to~2616~m~s$^{-2}$ with median (average) value of 57~m~s$^{-2}$ (34~m~s$^{-2}$) and it takes place up to a distance of about 28 R$_{SUN}$ with median (average) value of 7.8 R$_{SUN}$ (6 R$_{SUN}$). Additionally, the initial acceleration is significant in the case of fast CMEs (V$>$900~km~s$^{-1}$), where the median (average) values are about 295~m~s$^{-2}$ (251~m~s$^{-2}$), respectively, and much weaker in the case of slow CMEs (V$<$250~km~s$^{-1}$), where the median (average) values are about 18~m~s$^{-2}$ (17~m~s$^{-2}$), respectively. We note that the significant driving force (Lorentz force) can operate up to a distance of 6~R$_{SUN}$ from the Sun during the first 2 hours of propagation.
   We found a significant anti-correlation between the initial acceleration magnitude and the acceleration duration, whereas the residual acceleration covers a range  from $-$1224~to~0~m~s$^{-2}$ with a median (average) value of $-$34~m~s$^{-2}$ ($-$17~m~s$^{-2}$). One intriguing finding is that the residual acceleration is much smaller during the 24th cycle in comparison to the 23rd cycle of solar activity.
        Our study has also revealed that the considered parameters, initial acceleration (ACC$_{INI}$), residual acceleration (ACC$_{RES}$), maximum velocity (V$_{MAX}$), and  time at maximum velocity (Time$_{MAX}$) mostly follow solar cycles and the intensities of the individual cycle.}

\maketitle



%

\section{Introduction}

Space weather is mostly controlled by coronal mass ejections (CMEs), which are huge expulsions of magnetized plasma from the solar atmosphere. They have been intensively studied for their significant impact on the Earth's environment (e.g., \citealt{Bothmer07}, \citealt{Gosling91}, \citealt{Gopalswamy08}).
 The first CME was recorded by the coronograph on board the \emph{7th Orbiting Solar Observatory} (OSO-7) satellite (\citealt{Tousey73}). Since 1995 CMEs have been intensively studied using the sensitive \emph{Large Angle and Spectrometric Coronagraph} (LASCO) instrument on board the  \emph{Solar and Heliospheric Observatory} (SOHO) spacecraft (\citealt{Brueckner95}). SOHO/LASCO recorded about 30,000 CMEs until December 2017. The basic attributes of CMEs, determined manually from LASCO images, are stored in the SOHO/LASCO catalog\footnote{cdaw.gsfc.nasa.gov/CME$\_$list} (\citealt{Yashiro04}; \citealt{Gopalswamy09}). The initial velocity of CMEs, obtained by fitting a straight line to the height-time data points, has been the basic parameter used in prediction of geoeffectiveness of CMEs (\citealt{Manoharan04}, \citealt{Ravishankar19}).  Among the thousands of CMEs observed by LASCO coronagraphs from 1996 to
2017, only a few have speeds exceeding 3000~km~s$^{-1}$; however, an average CME speed tends to be on the order of 450~km~s$^{-1}$ (\citealt{Yashiro04}; \citealt{Webb12}). The yearly average CME speed changes with the solar cycle (\citealt{Yashiro04}) from $\approx$ 300 km~s$^{-1}$ during the minimum to $\approx$ 500 km~s$^{-1}$ during the maximum of solar activity. The rate of expansion of CMEs mainly depends on the Lorentz force that drives them and the conditions prevailing in the interplanetary medium (\citealt{Chen96}, \citealt{Kumar96}, \citealt{Kliem06}). In addition, the expansion is caused by the speed differences between the leading and trailing parts of the CME (\citealt{Chen00}). During the initial phase in the inner corona, the propelling Lorentz force is significant and ejections are accelerated rapidly. Farther from the Sun, in the outer corona, the magnetic force weakens and friction begins to dominate (\citealt{Sachdeva15}). The ejection speed drops, gradually approaching the speed of the solar wind. This smooth expansion can be disturbed by CME-CME interactions which can rapidly change velocities of ejections (\citealt{Gopalswamy01} \citealt{Gopalswamy02}). These collisions mostly occur during the maxima of solar activity. From the presented scenario, it appears  that a CME's kinematic evolution may undergo two distinct phases: an acceleration phase which is manifested by a rapid increase in speed and a propagation phase with almost constant velocity of propagation. Detailed research has shown that at the beginning of the acceleration phase we observe a slow increase in speed. This is the initiation phase of the ejection \citep{Zhang01, Zhang04}.

The initial phase of acceleration occurs in the inner corona ($\leq$5~R$_{SUN}$), while the second phase is at a greater distance ($\geq$5~R$_{SUN}$, \citealt{Zhang04}). The distribution of acceleration in the outer corona tend to be centered around zero with no significant variation of about $\pm$30~m~s$^{-2}$ (\citealt{Yashiro04}). Hence, this phase is also called  residual acceleration (\citealt{Chen03}, \citealt{Zhang06}). The situation is different in the case of acceleration in the inner corona. The initial or main acceleration can be diverse, ranging from extremely impulsive acceleration ($>$1000~m~s$^{-2}$) to extremely gradual acceleration ($<$20~m~s$^{-2}$),  (\citealt{Sheeley99}, \citealt{Zhang01}, \citealt{Zhang06}; \citealt{Alexander02}; \citealt{Gallagher03}).  However most of the CMEs have intermediate values of accelerations (\citealt{Wood99}; \citealt{Shanmugaraju03}; \citealt{Qiu04}, \citealt{Vrsnak04}; \citealt{Kundu04}; \citealt{Sterling05}; \citealt{Jing05}). \citet{Zhang06} made statistical study of the initial and the residual acceleration of 50 CMEs using all three LASCO coronagraphs (C1, C2, and C3) covering the field of view of about 1.1 - 32 R$_{SUN}$. Depending on the coronagraph and period of observation, the time cadence was about 12 - 30 minutes. They demonstrated that the initial acceleration can be in the range 2.8 - 4464~m~s$^{-2}$ with median (average) value of 170.1~m~s$^{-2}$ (330.9~m~s$^{-2}$), while the residual accelerations are in the range $-$131.0 to 52.0~m~s$^{-2}$ with median (average) value of 3.1~m~s$^{-2}$ (0.9~m~s$^{-2}$). It was shown that the duration of the main acceleration phase also covers a wide range of values of about 6 - 1200 minutes with median (average) value of 54 minutes (180 minutes). In addition, the magnitude of acceleration appears to be inversely proportional to the duration of acceleration. Unfortunately, this study was limited to small number of events and a short period of time (until mid-1998) when all three coronagraphs were functional. This study covered only a phase of minimum of solar activity.

The different phases of CME propagation described above are the result of interference of
 Lorentz force (propelling) and solar wind aerodynamic drag forces  (e.g., \citealt{Zhang06}; \citealt{Subramanian07}; \citealt{Gopalswamy13}). Other forces such as gravity or plasma pressure are  negligible in comparison to the Lorentz and drag forces for flux-rope models of CMEs (\citealt{Forbes00}; \citealt{Isenberg07}). In the first phase of propagation, up to a few solar radii, the Lorentz force prevails causing rapid CME acceleration  (e.g., \citealt{Vrsnak06}; \citealt{Bein11}; \citealt{Carley12}). This corresponds to the initial acceleration phase. At greater distances, friction begins to dominate, causing the CME to slow down, which corresponds to the residual acceleration phase. \citet{Sachdeva15}  show that for fast CMEs (speed $>$900~km~s$^{-1}$) the drag force starts to prevail from 5 R$_{SUN}$, but for slow CMEs (speed $<$900~km~s$^{-1}$) it is beyond 15 R$_{SUN}$. Recently, \citet{Sachdeva17}, using the graduated cylindrical shell (GCS) model,  consider the
 dynamics of a representative set of 38 CMEs observed with the SOHO and \emph{Solar and TErrestrial RElations Observatory} (STEREO) spacecrafts. They demonstrate that the Lorentz forces generally peak between 1.65 and 2.45~R$_{SUN}$ for the considered set of CMEs. They  also show that the Lorentz force becomes negligible around 3.5-4~R$_{SUN}$ for fast CMEs and beyond 12~R$_{SUN}$ for slow CMEs, respectively.

In this paper we present a statistical  study of the kinematic properties of 28894 CMEs recorded by  LASCO from 1996  to mid-2017. At the beginning of our research, the data in the catalog was available only until mid-2017. This research covers a large number of events observed during the 23rd and 24th solar cycles. For the study, we employed SOHO/LASCO catalog data and a new technique to determine the speed of ejections.

This paper is organized as follows. The data and method used for the study are described in Section 2. In Section 3 we present results of our study. Finally, the discussion and conclusions are presented in Section 4.

\section{Data and method}

The main aim of the study is to evaluate the acceleration and velocity of CMEs, up to $\approx$32~R$_{SUN}$ from the Sun in the LASCO field of view (LFV). For this purpose, observations from the SOHO/LASCO catalog  and  profiles of instantaneous velocities of CMEs were employed. It is worth mentioning that there are other CME catalogs mostly based on automatic measurements:  \emph{Computer Aided CME Tracking} (CACTus; \citealt{Robbrecht04}); \emph{Solar Eruptive Event Detection System} (SEEDS; \citealt{Olmedo08}); \emph{Automatic Recognition of Transient Events and Marseille Inventory from Synoptic maps} (ARTEMIS; \citealt{Boursier09}); \emph{CORonal IMage Processing} (CORIMP; \citealt{Byrne12});  \emph{Multi-View
CME} (MVC; \citealt{Vourlidas17}); and the FP7 \emph{Heliospheric Cataloguing, Analysis and Technique Service} (HELCATS; \citealt{Harrison16}), which  uses manual inspection of STEREO/HI data. In particular, the \emph{Coronal Mass Ejection Kinematic Database Catalogue} (KinCat; \citealt{Pluta19}), as part of the HELCATS catalog, has shown a similar study on a dataset of 122 CMEs to determine their initial speeds using GCS model.
 For our research, we decided to use the SOHO/LASCO catalog because the authors of this work are also contributors to this catalog. In our research we only use observations from the LASCO coronagraphs because, unlike the  STEREO observations, they cover up to two cycles of solar activity. In addition, the LFV is quite sufficient to study the effect of Lorentz and drag forces on CME propagation in Sun's environment (e.g., \citealt{Sachdeva15}).

Since 1996 CMEs have been routinely recorded by the sensitive LASCO coronagraphs on board the SOHO mission. The basic attributes of CMEs, determined manually from running 
difference images, are stored in the SOHO/LASCO  catalog. The two basic parameters, velocity and acceleration of CMEs, are obtained by fitting a straight and quadratic line to all
the height-time data points measured for a given  event. The parameters determined in this way, in some sense, reflect the average values in the field of view of the LASCO
coronagraphs. Nevertheless, it is evident that both these parameters are continuously changing with distance and time after CME onset from the Sun. 
Therefore, the average values of velocity and acceleration, used in many studies, do not give a correct description of CME propagation.

Initially, CME speed and acceleration increase rapidly when their dynamics are dominated by the Lorentz force. At the end of this phase of expansion, CMEs reach maximum velocities and their acceleration drops to zero. This means that forces acting on CMEs, at this point, are balanced. This first phase of  CME propagation  is  called  the initial or main acceleration. After reaching maximum speed,  the CMEs are gradually slowed down until they reach the speed of the solar wind. This phase of expansion is called    the residual acceleration. To recognize these phases of propagation we need to determine the instantaneous speeds and acceleration of CMEs.  For this purpose we used a new technique. To determine instantaneous velocities, we employed height-time measurements included in the SOHO/LASCO catalog. These height-time points are determined manually  at the fastest section of a given CME. Instantaneous ejection speeds are obtained from a linear fit to only five successive height-time data points.  This allows us to determine the instantaneous velocity at a given distance from the Sun.  Shifting such linear fit successively, by all height-time data points measured for a given CME, we obtain values of instantaneous velocities versus time or distance from the Sun. In other words, using a linear fit (for a very limited number of successive height-time points) we numerically determine  the derivative from the CME trajectory (i.e., we determine the instantaneous velocity). Formally we can obtain the instantaneous velocity using only two successive height-time points. Unfortunately, the manual measurements of height-time points are subject to unpredictable random errors. According to \citet{Michalek17}, the errors in determining velocity using linear regression are not significant.
The errors significantly depend on the number of height-time points measured for a given CME. For CMEs with ten or more height-time points, the velocity errors are always smaller than 15\% of the velocity itself and for five height-time points, the maximum errors can be about 20\%. In the case of acceleration, we may expect reasonable errors only for CMEs with more than 40 height-time measurements. After conducting several numerical tests, we decided to apply the technique described above (i.e., the five-point linear fit).

   \begin{figure}[h!]
\includegraphics[width=9.5cm,height=8cm]{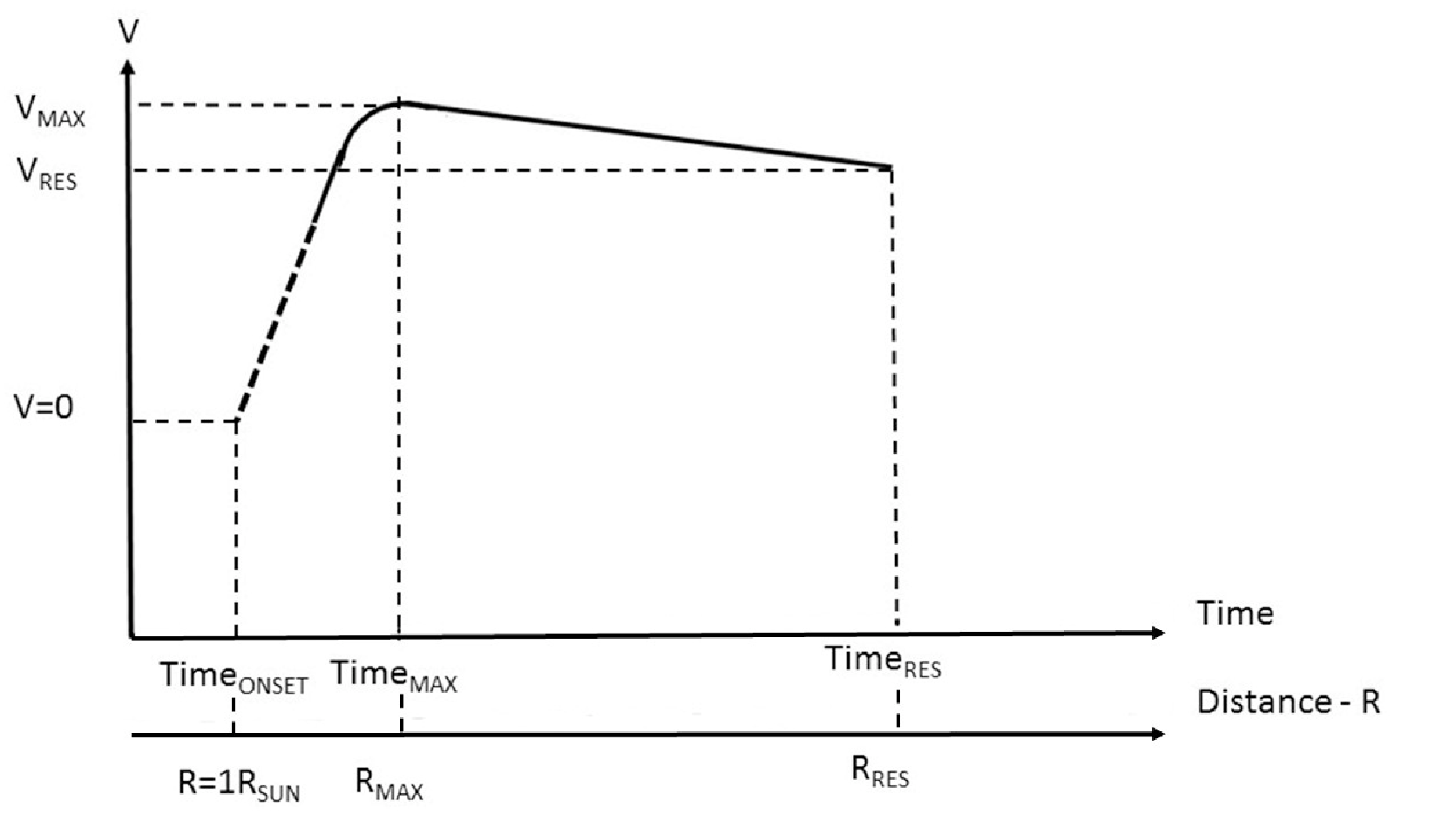}
  \caption{Schematic showing all the parameters introduced in this study. The continuous curve  represents the characteristic shape of velocities in time and distance. The thick dashed line illustrates the extrapolation of this line to the solar surface at distance, R=1~R$_{SUN}$ and velocity, V=0 km~s$^{-1}$. (R$_{MAX}$, R$_{RES}$) and (Time$_{MAX}$, Time$_{RES}$) are distance and time at maximum velocity (V$_{MAX}$) and residual velocity (V$_{RES}$), respectively.}
             \label{FigGam}%
  \end{figure}
 This technique allows us to obtain smooth profiles of instantaneous speeds versus time and distance from the Sun. Furthermore, these profiles can simply be used to obtain the maximum  (V$_{MAX}$) and residual (V$_{RES}$) velocities of CMEs. Having these profiles, we can also estimate times (Time$_{MAX}$, Time$_{RES}$) and distances (R$_{MAX}$, R$_{RES}$) when the V$_{MAX}$ and V$_{RES}$ are reached for a given event. It should be  noted that V$_{RES}$ is the final  velocity determined from  the profiles of instantaneous speed. These profiles are also used to estimate the initial and residual acceleration of CMEs. Figure~1 schematically shows all parameters introduced in this study. The continuous curve represents the characteristic shape (profile) of the instantaneous velocities in time and distance. The thick dashed line illustrates the extrapolation of this line to the solar surface (distance, R= 1~R$_{SUN}$ and velocity, V= 0~km~s$^{-1}$). The initial or main acceleration is obtained from the formula
$$
ACC_{INI}={{V_{MAX}}\over{ Time_{MAX}-Time_{ONSET}}},
$$
where $V_{MAX}$ is the maximum velocity of a given CME, $Time_{MAX}$ is the time at maximum velocity, and $Time_{ONSET}$ is the onset time of a given CME on the Sun. 
The onset time is defined here as the time obtained at V= 0 by backward extrapolation of the linear fit on the height-time plot to the solar surface (distance, R= 1~R$_{SUN}$). We note that these extrapolations are accurate only for limb events. For disk events the estimated onset is likely to be after the actual onset. The onset time determined by this method may be different from extreme UV observations and flare eruption times. Depending on an event (accelerating or decelerating) our onset time can be before or after the onset of associated flare. This difference depends on the acceleration of CMEs. Fortunately, the average acceleration of CMEs is close to zero (\citealt{Yashiro04}), so this effect does not affect our studies.
The residual acceleration can be obtained using the formula
$$
ACC_{RESIDUAL}={{V_{RES}-V_{MAX}} \over {Time_{RES}-Time_{MAX}}},
$$
where $V_{RES}$ and $Time_{RES}$ are the last determined velocity and time for a given event, respectively. This method was employed to the 28894 CMEs included in the SOHO/LASCO catalog in the considered period, 1996 - 2017. It is worth noting that this method can be used if the CME has at least eight height-time data points. Hence, we could not use it for very fast events (V$>$2000~km s$^{-1}$). In the considered period the catalog included 28894 events. For 21492 events we were able to determine the velocity profiles. This means that 7402 events had fewer than nine height-time points to obtain reasonable speed profiles (including 48 very fast events which were excluded from further consideration). Using these new parameters and data included in the SOHO/LASCO catalog, we conducted statistical analysis on kinematics of CMEs during the 23rd and 24th solar cycles. However, we note a certain shortcoming of the proposed research. Unfortunately, the C1 coronagraph that was used to observe near the surface of the Sun has not been operational since the temporary loss of SOHO in June 1998. In our research we used images from the C2 and C3 coronagraphs whose field of view covers the area 2 - 32 R$_{SUN}$. This means that we do not observe the area near the Sun where the initial phase of acceleration of many CMEs can occur. Therefore, the initial accelerations determined using our method should be formally considered as lower limits. Fortunately, this inaccuracy does not have a significant impact on our considerations because in our study we focus on observing general trends in the kinematics of CMEs during the last two cycles of solar activity.

 It is also important to note that coronagraphic observations of CMEs are subject to projection effects. This makes it practically impossible to determine the true velocities or accelerations of CMEs.  \citet{Bronarska18} demonstrated that this effect depends significantly on width and source location of CMEs. We cannot eliminate it in our statistical analysis. Therefore, we note that the CME parameters determined  in this study may differ from the real values. They can be considered  an underestimation.

\section{Statistical results}
In our study, we considered 28894 CMEs recorded by the SOHO/LASCO coronagraph from 1996 to mid-2017. At the time we started our study the catalog contained observations only until mid-2017. Using our method we estimated the introduced parameters characterizing  the kinematics of CMEs in different phases of their expansion. We were able to determine ACC$_{INI}$ and ACC$_{RES}$ for 21492 (74$\%$) and 17092 (60$\%$) events, respectively.

  \subsection{Acceleration of CMEs}
  
\begin{figure*}[!h]
  \centering
  \includegraphics[width=18cm,height=8cm]{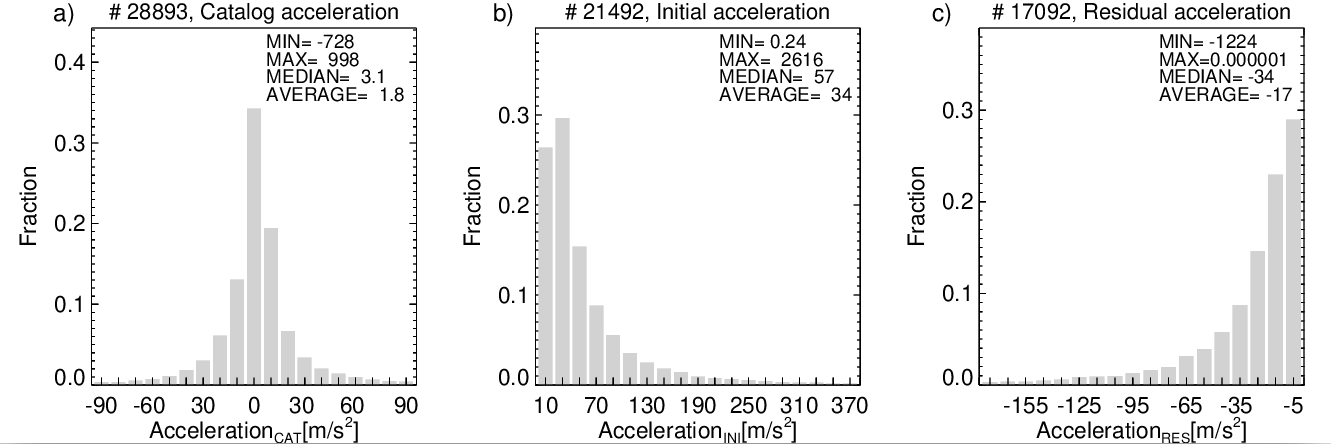}
   \caption{Distributions of the catalog (panel~a), initial (panel~b), and residual (panel~c) acceleration of 28894 CMEs in the SOHO/LASCO catalog during the period 1996 - 2017. In the upper right corners the minimum, maximum, median, and average values are displayed.}
             \label{FigGam}%
   \end{figure*}

In the previous sections we present the method for determining the initial and residual accelerations that are observed during the beginning phase of CME propagation near the Sun, in the LFV. In this section we compare  these parameters with the acceleration determined for the SOHO/LASCO catalog. In Figure 2 the successive panels show distributions of the catalog, initial, and residual acceleration. We note that the distribution of these accelerations are completely different. As we could expect (\citet{Yashiro04}),  the catalog acceleration (panel~a) has a normal distribution with the peak near zero. This means that on average CMEs in the field of view of LASCO coronagraphs do not show acceleration. However, detailed analysis (\citet{Yashiro04}) reveals  that slow CMEs (V$<$250~km~s$^{-1}$) are mostly accelerated and fast CMEs (V$>$900~km~s$^{-1}$) are mostly decelerated. This behavior is due to the interaction of the CME with the surrounding solar wind. The  drag force slows down all the ejections faster than the solar wind and speeds up the ones that are slower than solar wind (\citealt{Michalek15}). 
This shows  that the drag force  could be very important for  CME propagation up to 32 R$_{SUN}$. The catalog acceleration is the result of two acceleration phases: the initial and residual acceleration. The initial acceleration has a wide distribution from 0.24 to 2616~m~s$^{-2}$, with median and average values equal to 57 and 34~m~s$^{-2}$, respectively. However, 70\% of the considered CMEs have no significant acceleration during the initial phase of propagation ($\leq$60~m~s$^{-2}$). Only a small percentage of CMEs were subject to significant acceleration ($\geq$250~m~s$^{-2}$).
A completely different result is observed for the residual acceleration. The residual acceleration has a distribution from 1224~to~0~m~s$^{-2}$ with median and average values equal to $-$34 and $-$17~m~s$^{-2}$, respectively. More than 70\% of the considered CMEs have insignificant deceleration during the residual phase of propagation ($\geq-$30~m~s$^{-2}$). Only a small percentage of CMEs were subject to notable deceleration ($\leq-$125~m~s$^{-2}$). The  obtained results show that the kinematics of CMEs in the LFV is more complex than appears from the data included in the SOHO/LASCO catalog.

   \begin{figure*}[!h]
  \centering
  \includegraphics[width=18cm,height=9cm]{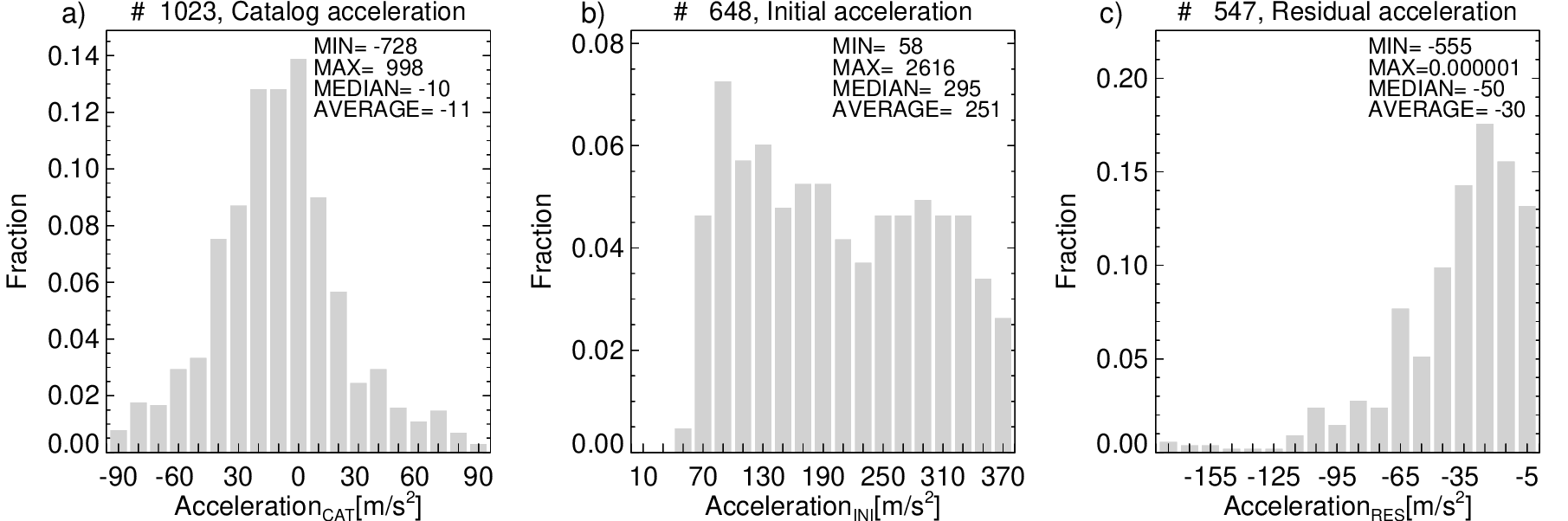}
   \caption{Distributions of the catalog (panel~a), initial (panel~b), and residual (panel~c) acceleration of the fast CMEs (V$>$900~km~s$^{-1}$) in the SOHO/LASCO catalog. In the upper right corners the minimum, maximum, median, and average values are displayed.}
             \label{FigGam}%
   \end{figure*}

 \begin{figure*}[!h]
  \centering
  \includegraphics[width=18cm,height=9cm]{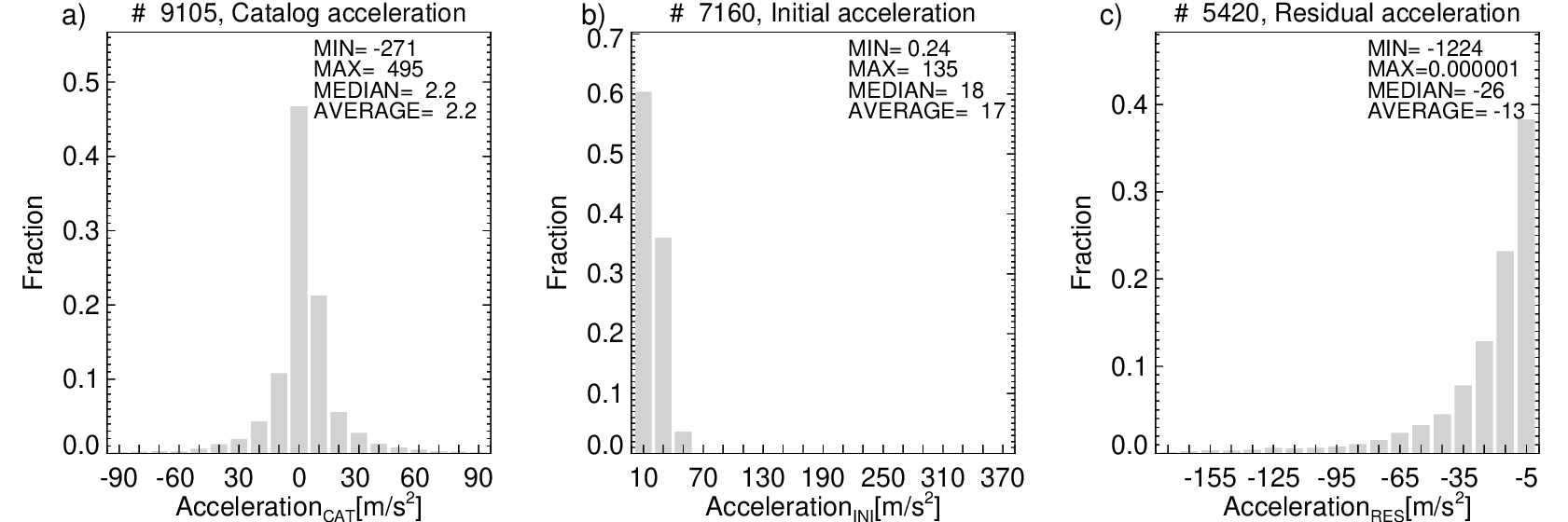}
  \caption{Distributions of the catalog (panel~a), initial (panel~b), and residual (panel~c) acceleration of the slow CMEs (V$<$250~km~s$^{-1}$) in the SOHO/LASCO catalog. In the upper right corners the minimum, maximum, median, and average values are displayed.}
            \label{FigGam}%
   \end{figure*}

Next, in Figure~3~and~4 the same distributions are shown, but for fast (V$>$900~km~s$^{-1}$) and slow (V$<$250~km~s$^{-1}$) CMEs, respectively.
These figures show that the fast events are, on average, decelerated (Figure~3, panel~a, SOHO/LASCO catalog acceleration). However, they are subject to significant acceleration during the initial phase (Figure~3, panel~b, median = 295~m~s$^{-2}$ and average = 251~m~s$^{-2}$) and deceleration during the residual phase (Figure~3, panel~c, median = $-$50~m~s$^{-2}$ and average = $-$30~m~s$^{-2}$). Although the residual acceleration is on average less significant  than the initial acceleration, it lasts longer and therefore the average (catalog) acceleration for these CMEs is negative.
  The opposite conclusion can be drawn for the slow events. They are on average slightly accelerated (Figure~4, panel a). However, they are subject to insignificant acceleration during  the initial phase (Figure~4, panel b, median = 18~m~s$^{-2}$ and average = 17~m~s$^{-2}$) and deceleration during the residual phase of evolution (Figure~4, panel c, median = $-$26~m~s$^{-2}$ and average = $-$13~m~s$^{-2}$).
  Although the initial acceleration is, on average, less significant  than the residual acceleration, it lasts longer and therefore the average (catalog) acceleration for these CMEs is positive.


   \begin{figure*}
  \centering
  \includegraphics[width=16cm,height=12cm]{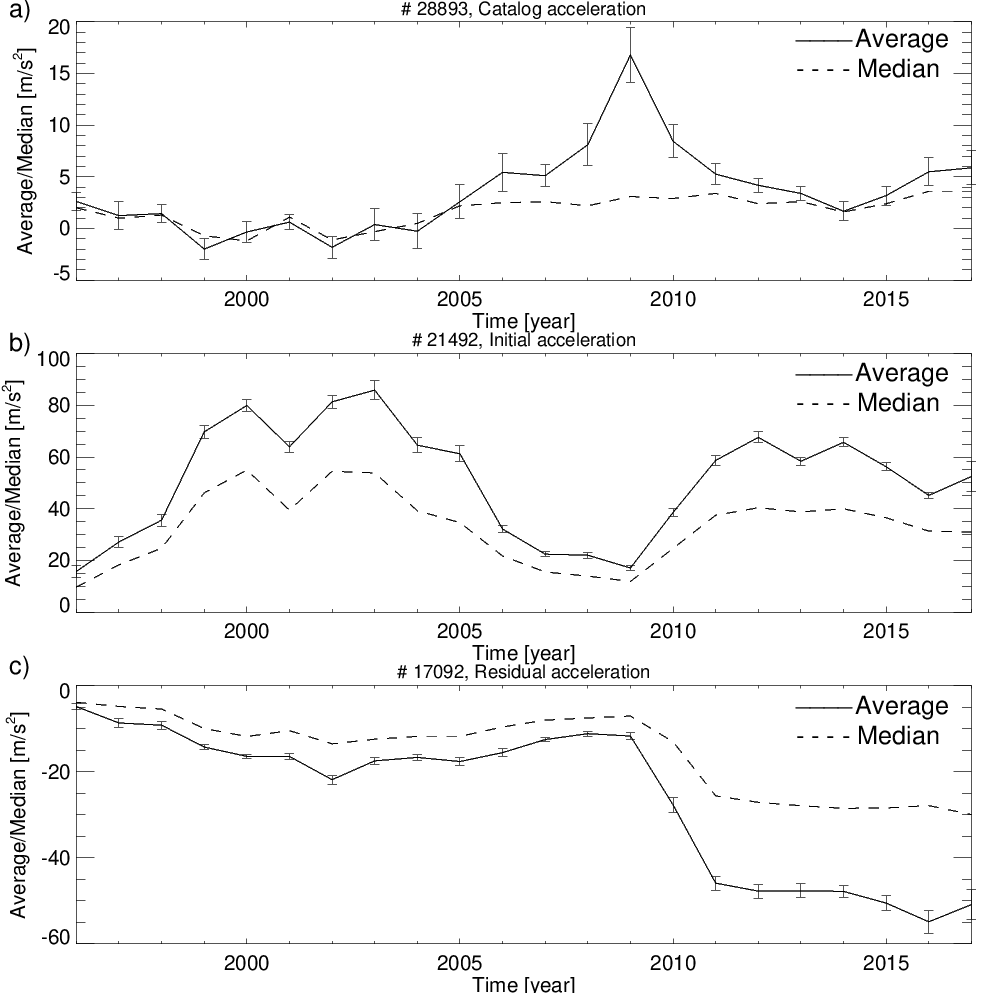}
  \caption{Yearly median and average values of the catalog (panel~a), initial (panel~b), and residual (panel~c) acceleration vs time of the CMEs in the SOHO/LASCO catalog. The error bars are standard errors on the average.}
             \label{FigGam}%
   \end{figure*}

It is also interesting to recognize the evolution of these parameters with solar cycle. In Figure~5 we show changes of the yearly median and average values of these parameters with solar cycles 23 and 24. From the figure it is clear that these parameters behave differently over time. In the case of the catalog acceleration there is no clear correlation with solar cycles (panel~a). During solar cycle 23, up to 2008, the average value of the catalog acceleration oscillates around zero. At the end of the cycle (in 2006), this parameter starts to increase, reaching its maximum value in 2009 ($\approx15~m~s^{-2}$). Then the average value drops, but never reaches 0~m~s$^{-2}$. This intriguing  behavior is caused by the extraordinary minimum of solar activity that appeared between the 23rd and 24th cycles. During this period, several weak and slow events were recorded. These events were driven not by the Lorentz force, but mainly by the drag force from the surrounding solar wind (\citealt{Michalek19}). For these events the drag force causes acceleration through almost the entire LFV.

\begin{figure*}[h!]
  \centering
  \includegraphics[width=16.5cm,height=7cm]{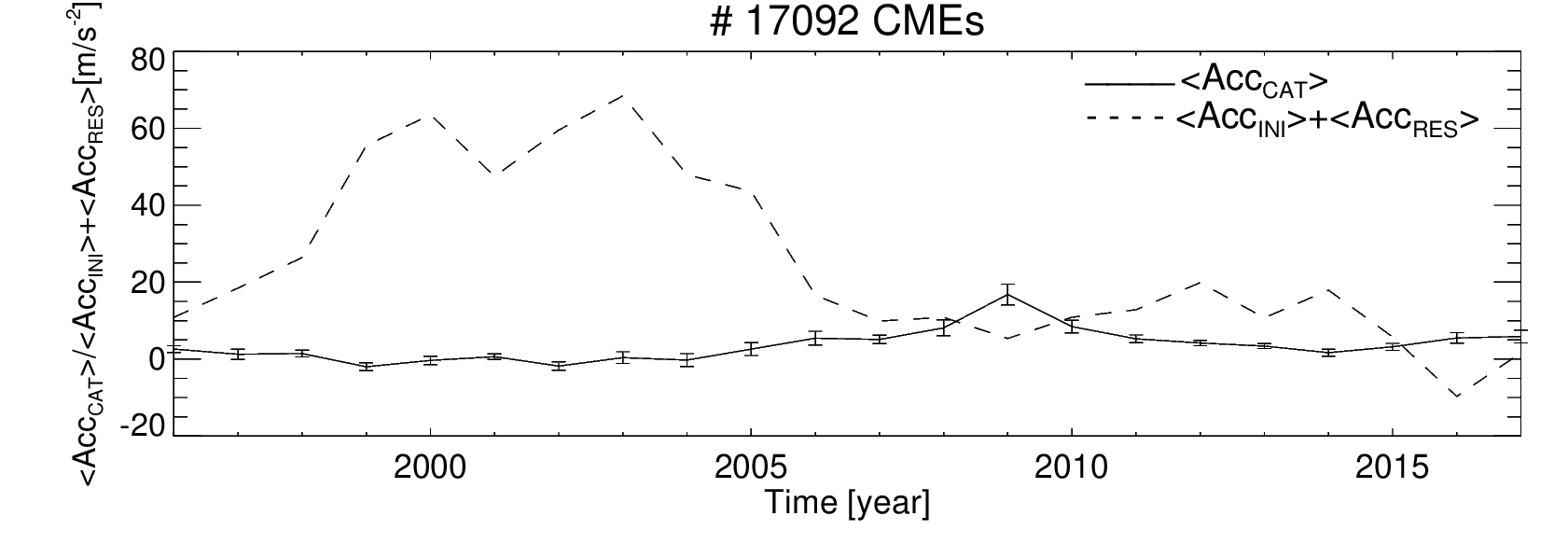}
  \caption{Yearly average value of catalog acceleration (solid line) vs time and a sum of the yearly average value of the initial and residual acceleration vs time (dashed line) of the CMEs in the SOHO/LASCO catalog.
The error bars are the standard error on the average.}
             \label{FigGam}%
   \end{figure*}

 In panel b we have the yearly median and average values of initial acceleration versus time. In this case we observe a clear relation of these parameters with solar cycles. During the solar maximum, on average the initial acceleration is much higher compared to the solar minimum. The average initial acceleration reaches a value of over 80~m~s$^{-2}$ and about 70~m~s$^{-2}$ during the maxima of cycles 23 and 24, respectively. During the minima of both cycles it is only about 20~m~s$^{-2}$. This result is consistent with our expected values;  during the maximum of solar activity the CMEs on average reach higher speeds, and therefore  must be subject to significantly higher acceleration. This is also consistent with the amplitude of both cycles determined on the basis of sun spot numbers (SSNs). In both cycles, we observe double peaks similar to those obtained using other parameters, like SSNs and CME numbers, indicating  solar activity.

 A very different behavior is observed in the case of  residual acceleration (panel~c). During solar minimum, the average value of residual accelerations is close to 0~m~s$^{-2}$ (1996 and 2009).
 During the maximum this parameter, on average, reaches a significant negative value of about $-$20~m~s$^{-2}$ and $-$50~m~s$^{-2}$ in cycles 23 and 24, respectively. This result is also consistent with our expected values. During the solar maximum on average the CMEs reach higher speeds and therefore must be in the residual phase when propelling force decays, leading to a significant deceleration.  It seems that the residual acceleration follows  solar cycles, but it is intriguing that during the maximum of cycle 24 we observe a sudden significant drop in the average residual acceleration. This is due to the anomalous behavior of CMEs in the 24th solar cycle. A study by \citet{Gopalswamy15} and \citet{Selvakumaran16} demonstrated that during this solar cycle CMEs were less energetic, but still expanded (in width) rapidly in the reduced total pressure (plasma + magnetic) of the interplanetary medium. This resulted in the quick decrease of the propelling force (Lorentz force) as the internal magnetic field diffused quickly. This led to the drag force prevailing significantly over the propelling force  compared to the previous maximum of solar activity. This reveals itself as a considerable drop in the residual acceleration of 24th cycle (panel~c). The LASCO coronagraph cadence was about 1 image per hour, but it is important to note that in 2010 (at the beginning of the 24th cycle) the LASCO coronagraph instrument image cadence was doubled. However, the conducted tests showed that this did not have a significant impact on our analysis.

It is also worth noting that the catalog acceleration is not a simple sum of the initial and residual accelerations.
 Figure~6 shows the yearly average value of the catalog acceleration and a sum of the yearly average values of the initial and residual acceleration versus time. The catalog acceleration  during the two solar cycles fluctuates around zero (with one exception in 2009). However, the combined average acceleration ($Acc_{INI}+Acc_{RES}$) follows the activity of solar cycles. It has two distinct peaks during the maxima,  at more than 60~m~s$^{-1}$ and about 20~m~s$^{-1}$ in the 23rd and 24th cycles, respectively. The amplitude of these peaks reflects the intensities of solar activity in the two cycles.  The combined average acceleration is significantly positive in the maximum of the 23rd cycle ($\approx$70~m~s$^{-1}$).  During the maximum of cycle 23, CMEs are powerful and are on average, significantly accelerated to greater distances from the Sun in comparison to events in the maximum of cycle 24.

The presented  statistical analysis revealed that at the beginning of their expansion, in the vicinity of the Sun, CMEs are subject to many factors (Lorentz Force, CME-CME interaction, speed differences between leading and trailing parts of the CME) that determine their propagation. Although their average values of the catalog accelerations are always close to zero, a more detailed study shows that their instantaneous accelerations may be quite different depending on the conditions prevailing in the Sun and the environment in which they propagate. These conditions vary depending on the individual eruption and over time as the solar activity changes.

\begin{figure*}
\begin{minipage}{22cm}
\includegraphics[width=6.3cm,height=6cm]{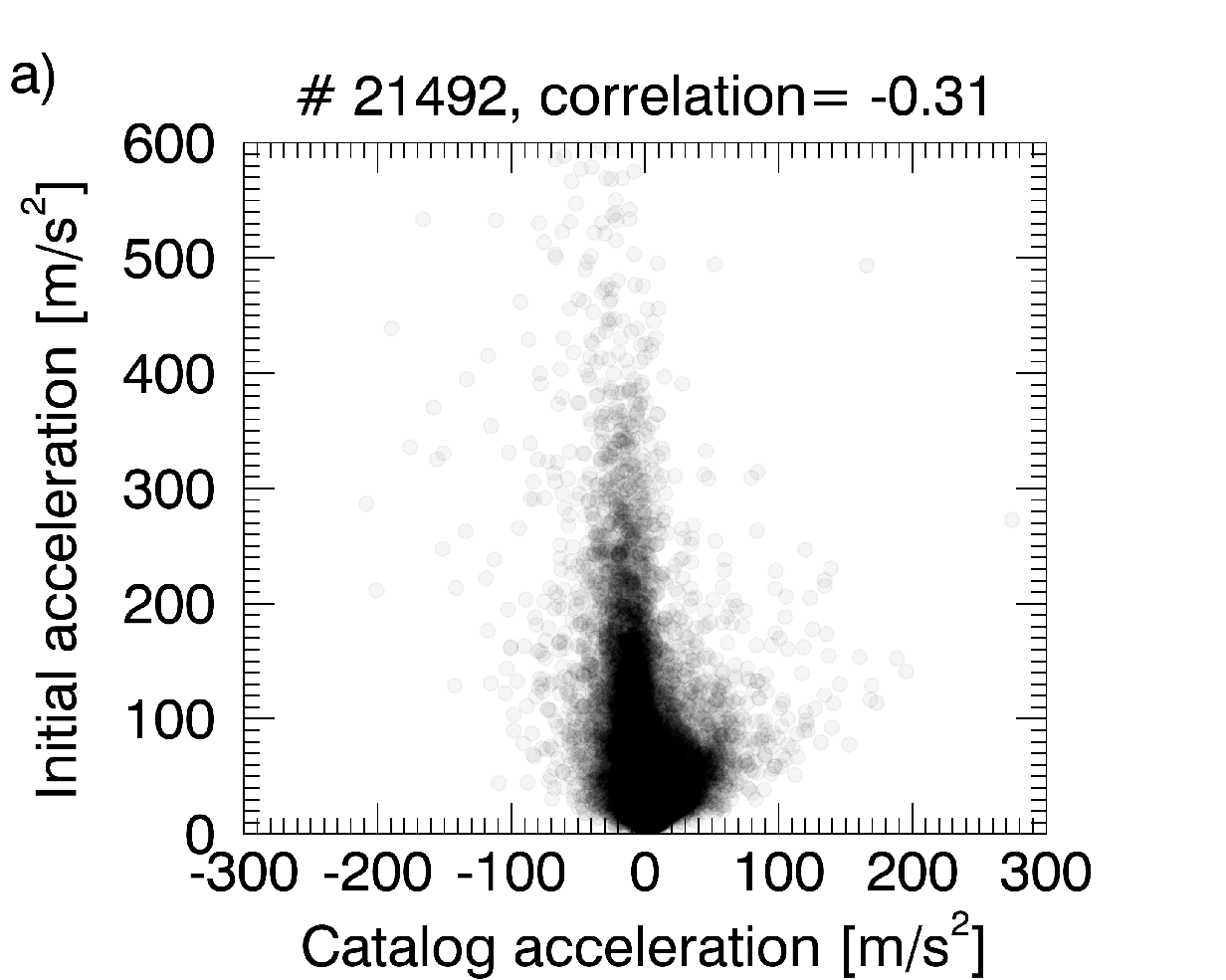}
\includegraphics[width=6.3cm,height=6cm]{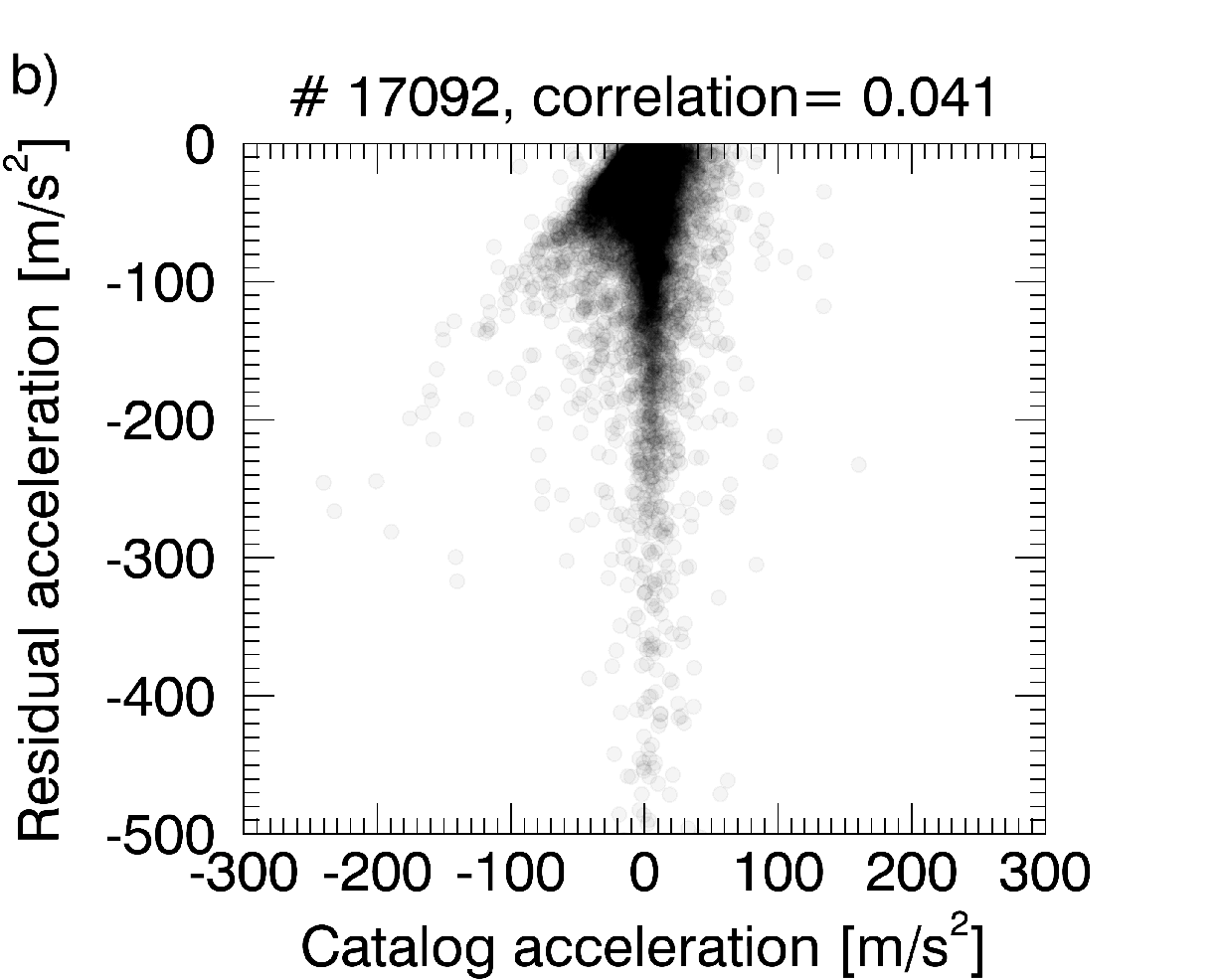}
\includegraphics[width=6.3cm,height=6cm]{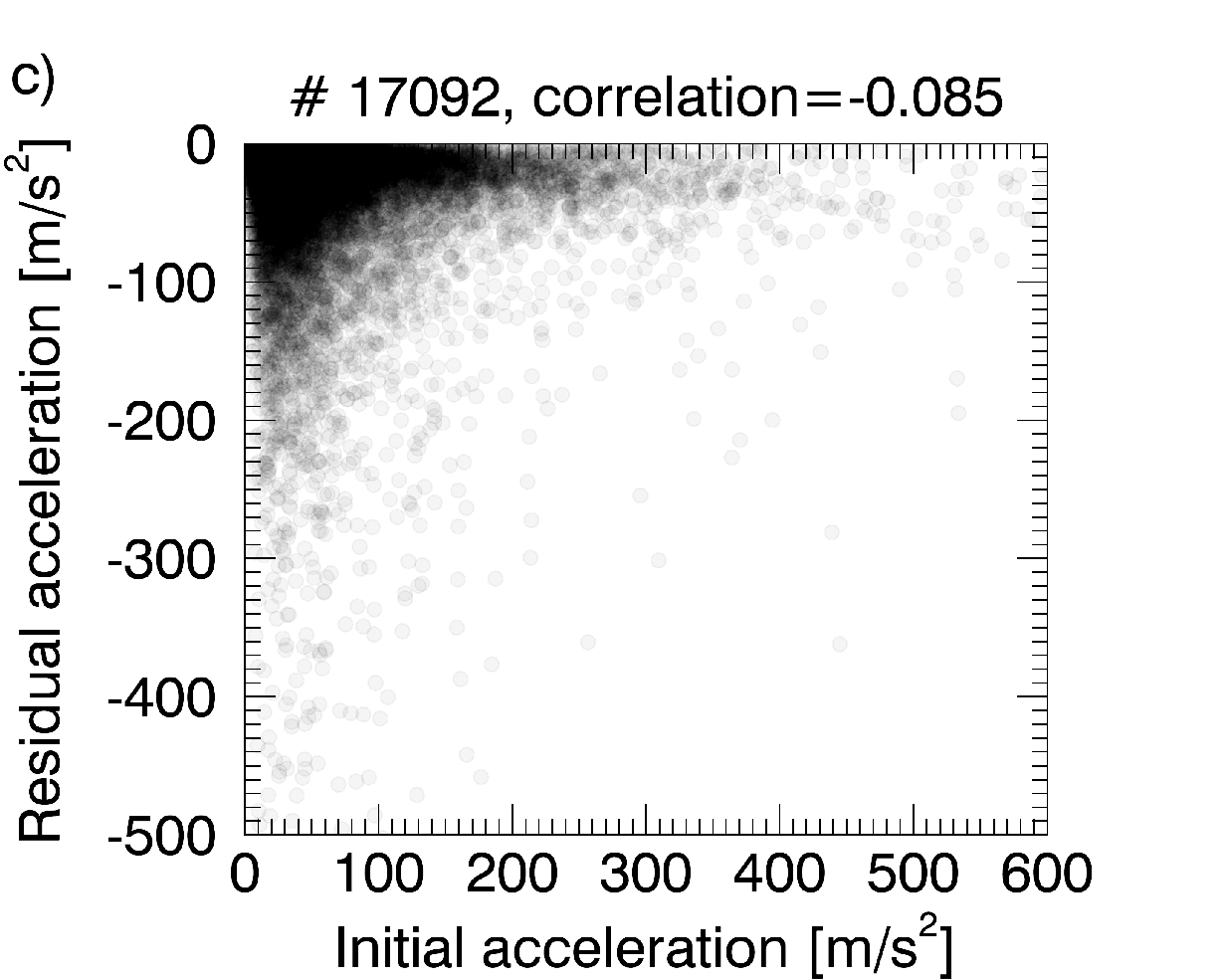}
\end{minipage}
\caption{Scatter plots of the catalog vs initial acceleration (panel a), the catalog vs residual acceleration (panel b), and the initial vs residual acceleration (panel c) of the CMEs in the SOHO/LASCO catalog.}
\end{figure*}

\begin{figure*}
  \centering
  \includegraphics[width=18cm,height=9cm]{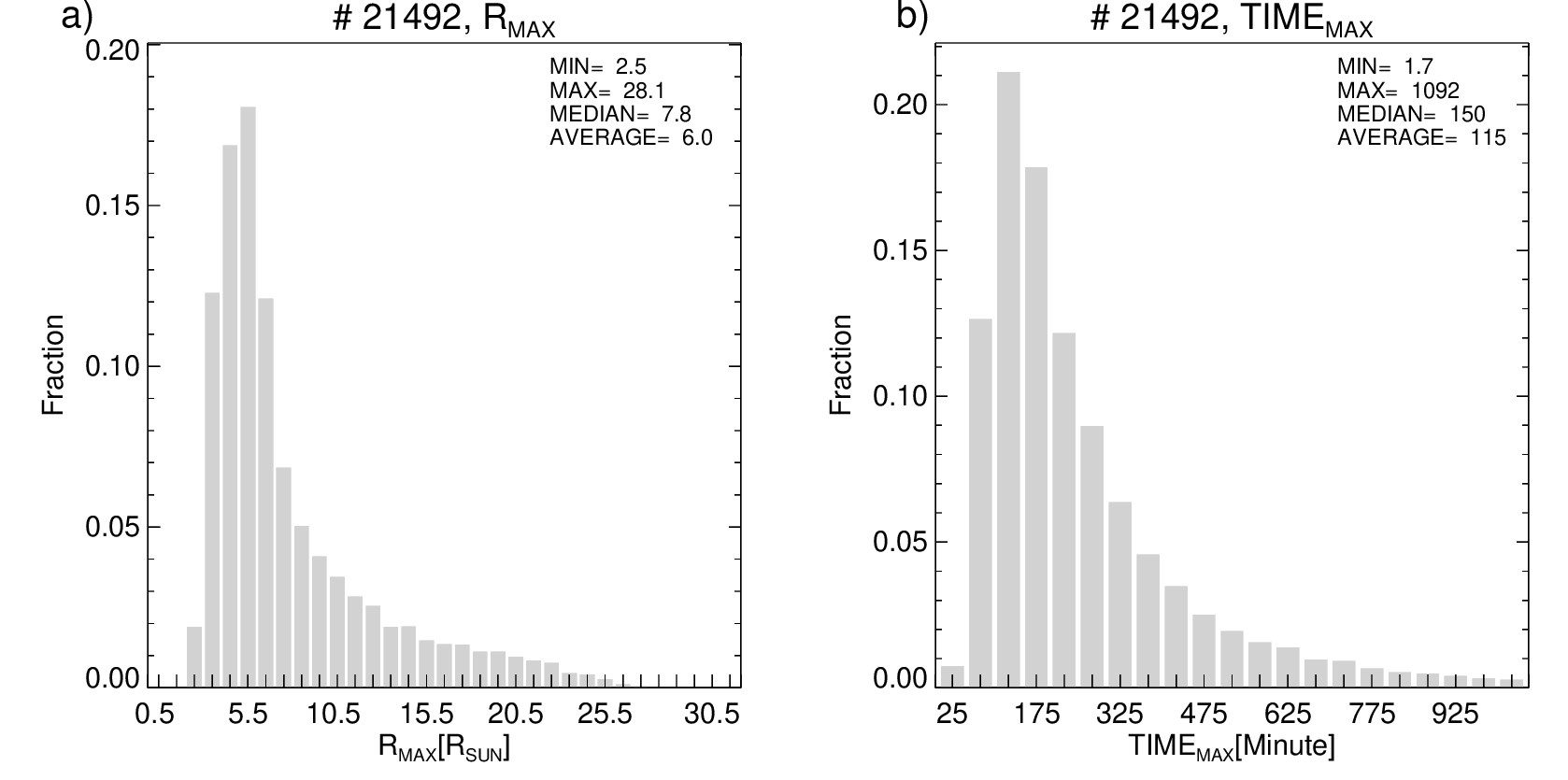}
  \caption{Distributions of distance (R$_{MAX}$, panel a) and time (Time$_{MAX}$, panel b) when CMEs reach  maximum velocity.}
             \label{FigGam}%
   \end{figure*}

\begin{figure*}[h!]
  \centering
  \includegraphics[width=16cm,height=10cm]{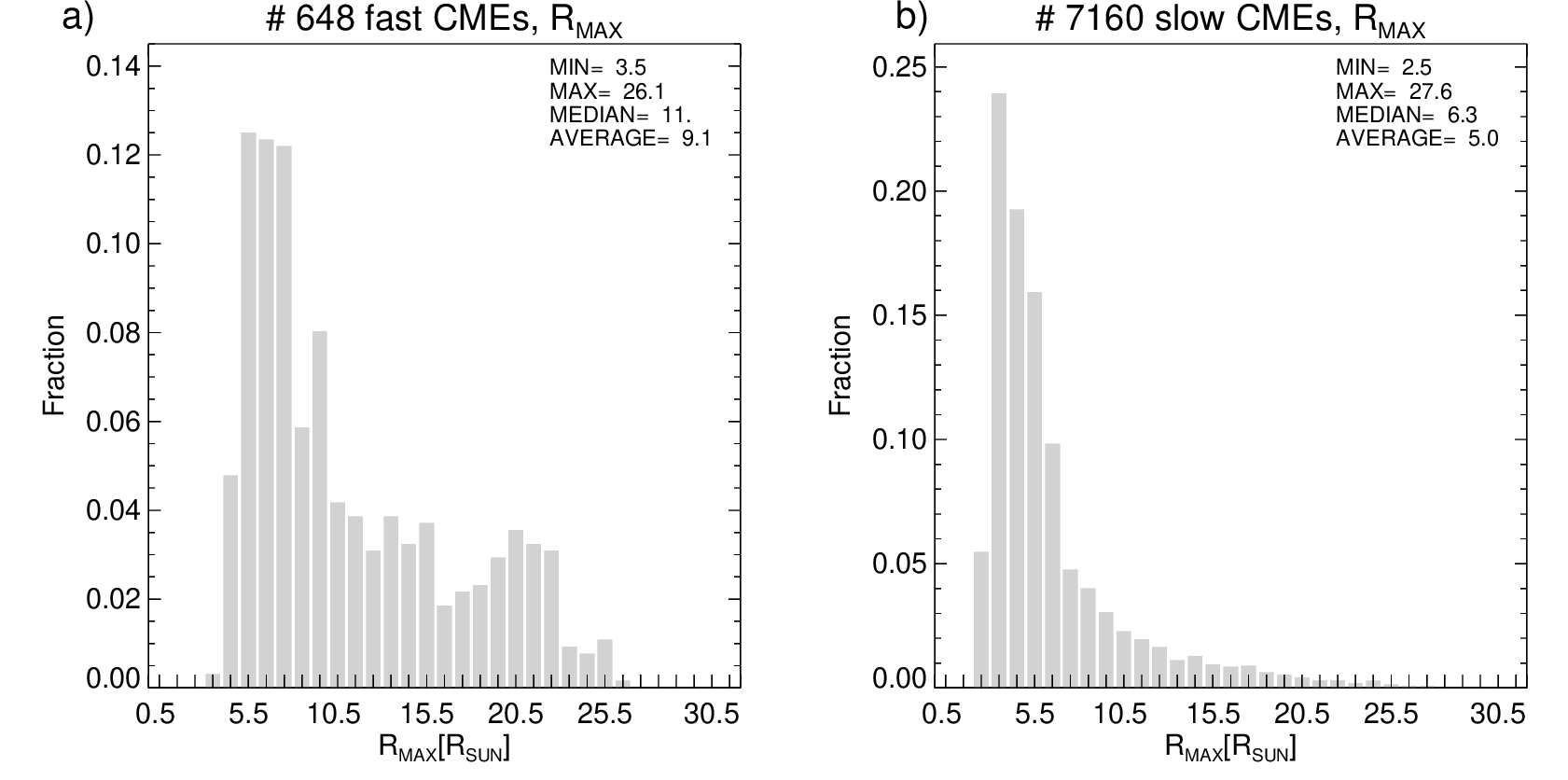}
  \caption{Distributions of the distance when CMEs reach the maximum velocity, R$_{MAX}$, for fast CMEs (V$>$900 km/s, panel a) and slow CMEs (V$<$250 km/s, panel b).}
             \label{FigGam}%
  \end{figure*}

\begin{figure*}
\begin{minipage}{14cm}
\includegraphics[width=7cm,height=6cm]{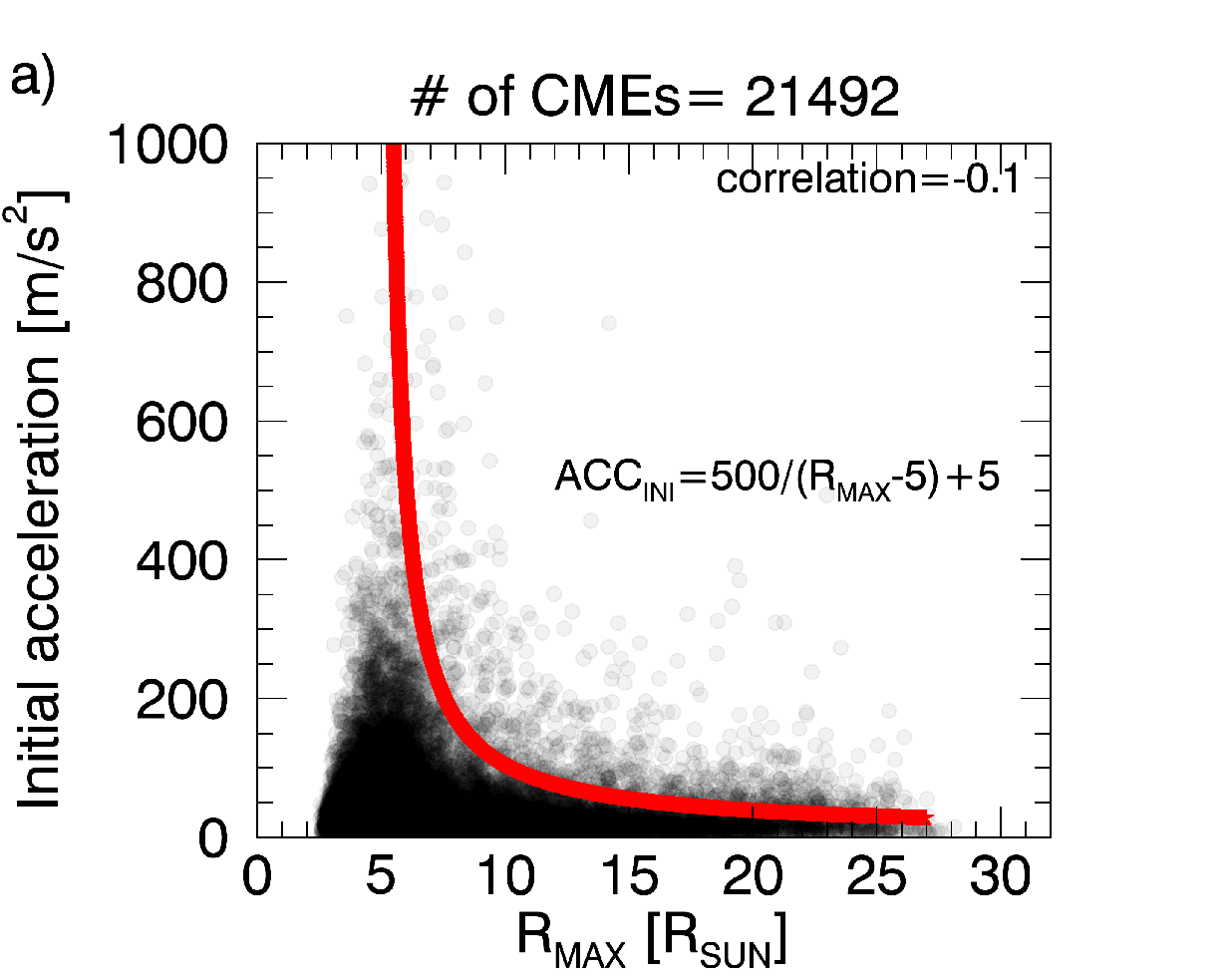}
\includegraphics[width=7cm,height=6cm]{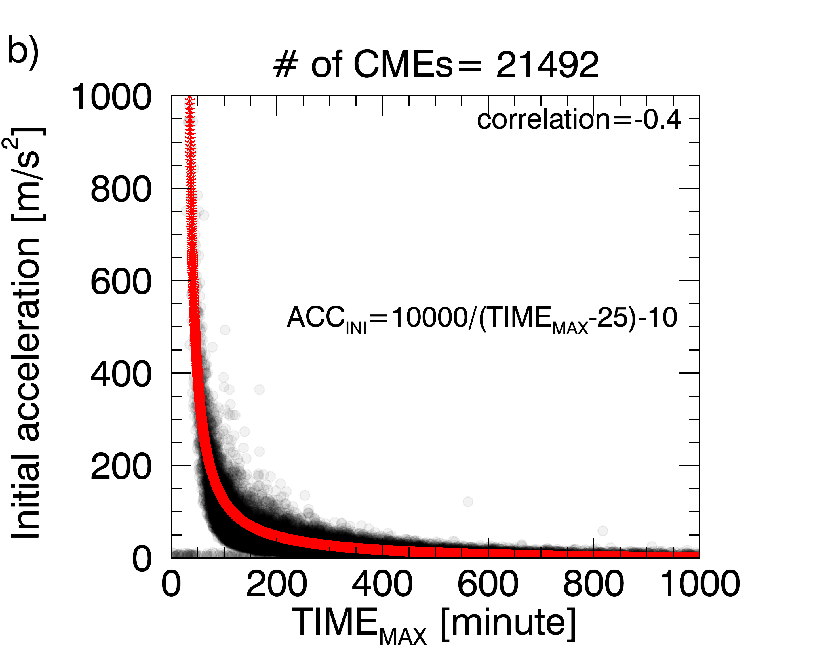}
\includegraphics[width=7cm,height=6cm]{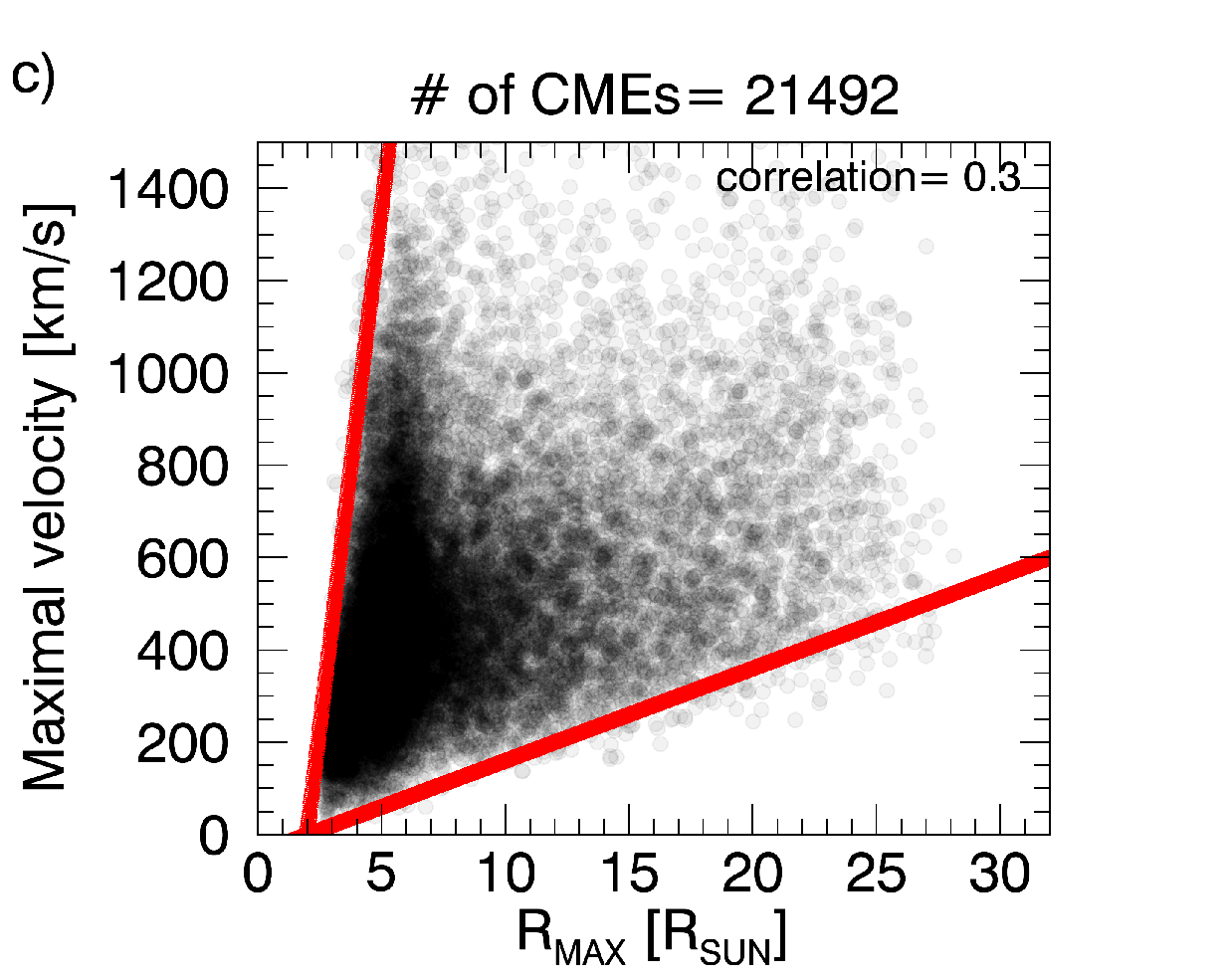}
\includegraphics[width=7cm,height=6cm]{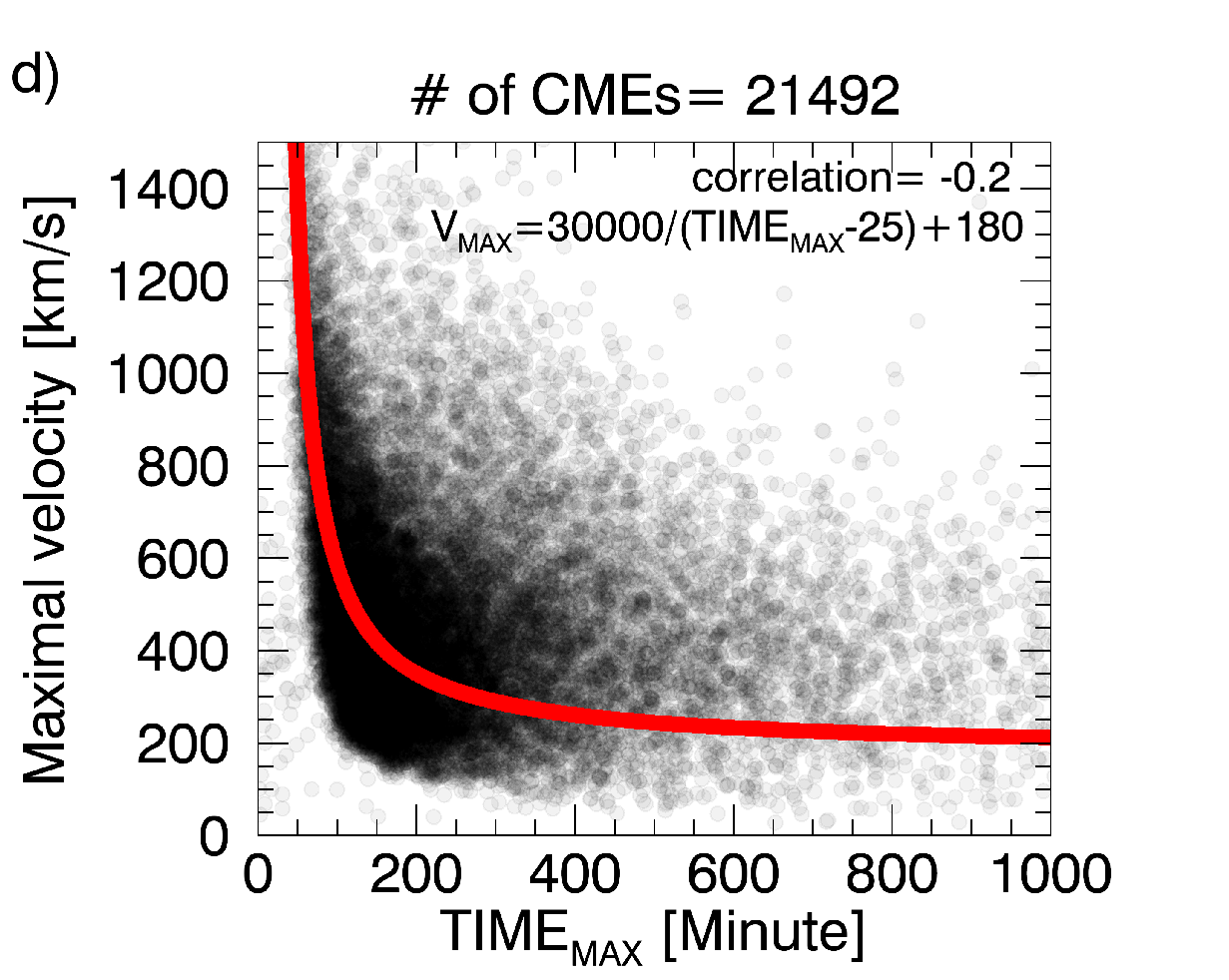}
\end{minipage}
\caption{Scatter plots of initial acceleration vs R$_{MAX}$ (panel a) and Time$_{MAX}$ (panel b), and the maximum velocity vs R$_{MAX}$ (panel c) and Time$_{MAX}$ (panel d).}

\end{figure*}

Finally in Fig.~7,  we present scatter plots showing the relations between the respective accelerations. We do not observe any significant correlation between the considered accelerations (correlation coefficients are less than 0.31). Nevertheless, it is worth  noting that the initial and residual accelerations (panel~c) are anti-correlated. The CMEs that have the strongest initial acceleration have low residual deceleration and vice versa.

\subsection{Range of initial acceleration}

Using our analysis we can determine the distance (R$_{MAX}$) and time (Time$_{MAX}$) when a given CME reaches maximum velocity (see Figure~1).
 In Figure~8 we present the distributions of  R$_{MAX}$ (panel a) and Time$_{MAX}$  (panel b).
The R$_{MAX}$ is in the range 2.5 - 28.1 R$_{SUN }$ with median (average) value of 7.8 R$_{SUN}$ (6.0 R$_{SUN }$). The Time$_{MAX}$ is in the range 1.7 - 1092~minutes with median (average) value of 150 minutes (115) minutes. The clearly distinguishable peaks in these distributions allow us to conclude that CMEs are mostly accelerated up to distance of $\approx5.5$ R$_{SUN}$ within about two hours of the onset. This is in contradiction with previous research, which showed that the main acceleration phase takes place over a distance of about 3.3 R$_{SUN}$ (\citealt{Zhang04}). However, these studies were based on a small population of events. We have  found a significant correlation (Pearson correlation=0.56), proved by a significance test of p=0.05, between the R$_{MAX}$ and Time$_{MAX}$.

Figure~9 shows the distribution of R$_{MAX}$ for fast events (V$>$900 km s$^{-1}$, panel~a) and slow events (V$<$250 km s$^{-1}$, panel~b). Since the distributions are different it proves   that the R$_{MAX}$ depends on the velocity of CMEs. Only slow events can be accelerated close to the Sun, within the distance $\geq$2.5 R$_{SUN}$  of the Sun (this limit is due to the LASCO C2 field of view), with median and average values of about 6.3 and 5.0 R$_{SUN}$, respectively. A large number of these slow events are accelerated only to 10~R$_{SUN}$ (panel~a). The faster events need a greater acceleration distance to achieve the maximum speed. The minimum acceleration distance for these CMEs is about 3.5 R$_{SUN}$ (average and median values are 11.0 and 9.1 R$_{SUN}$, respectively). A large number of these fast events are accelerated up to the distance 20~R$_{SUN}$ (panel~b). The findings from this paper differ from those of the previous study by \citet{Sachdeva15, Sachdeva17} as there are a greater  number of events considered in our analysis. These trends are also presented in Figure~10 (panel~c).

Figure~10 shows, for all the considered CMEs, scatter plots of Acc$_{INI}$ versus R$_{MAX}$ (panel~a), Acc$_{INI}$ versus Time$_{MAX}$ (panel~b), V$_{MAX}$ versus R$_{MAX}$ (panel~c), and V$_{MAX}$ versus Time$_{MAX}$ (panel~d).  Although there is no significant correlation between these parameters, we can observe some general trends. From panel~a we see that the strong initial acceleration (>100~m~s$^{-1}$) can only occur over the distance of about 6 R$_{SUN}$ from the Sun. The approximate reflection of this trend is shown by the red line. This line is described by the equation presented in the panel. Panel b shows a clear hyperbolic relation between Acc$_{INI}$ and Time$_{MAX}$, presented by the red line in the panel. It is in agreement with the previous study by \citet{Zhang06}. This relation also shows that the strong initial acceleration (>100~m~s$^{-1}$) can only occur within two hours of the CME onset. It should be clearly emphasized that the significant driving force (Lorentz force) can only operate up to a distance of 6~R$_{SUN}$ from the Sun and during the first two hours of propagation.

Panel~c shows the V$_{MAX}$  versus R$_{MAX}$. There is no clear correlation between these parameters; however, we observe a few characteristic trends. First of all, it is clear that only slow CMEs can have a short distance of the initial acceleration. The faster CMEs are accelerated over larger distances. These trends are presented using a red line (the line with a higher inclination). Secondly, the greater the acceleration distance, the higher the speed of the CMEs. These trends are also shown in red (the line with the lower inclination).  The last panel~d presents the V$_{MAX}$ versus Time$_{MAX}$. The results are similar to  the case of the initial acceleration; we observe an anti-correlation between these parameters. It is described by the hyperbolic relationship shown in red.

\begin{figure*}
  \centering
  \includegraphics[width=14cm,height=8.1cm]{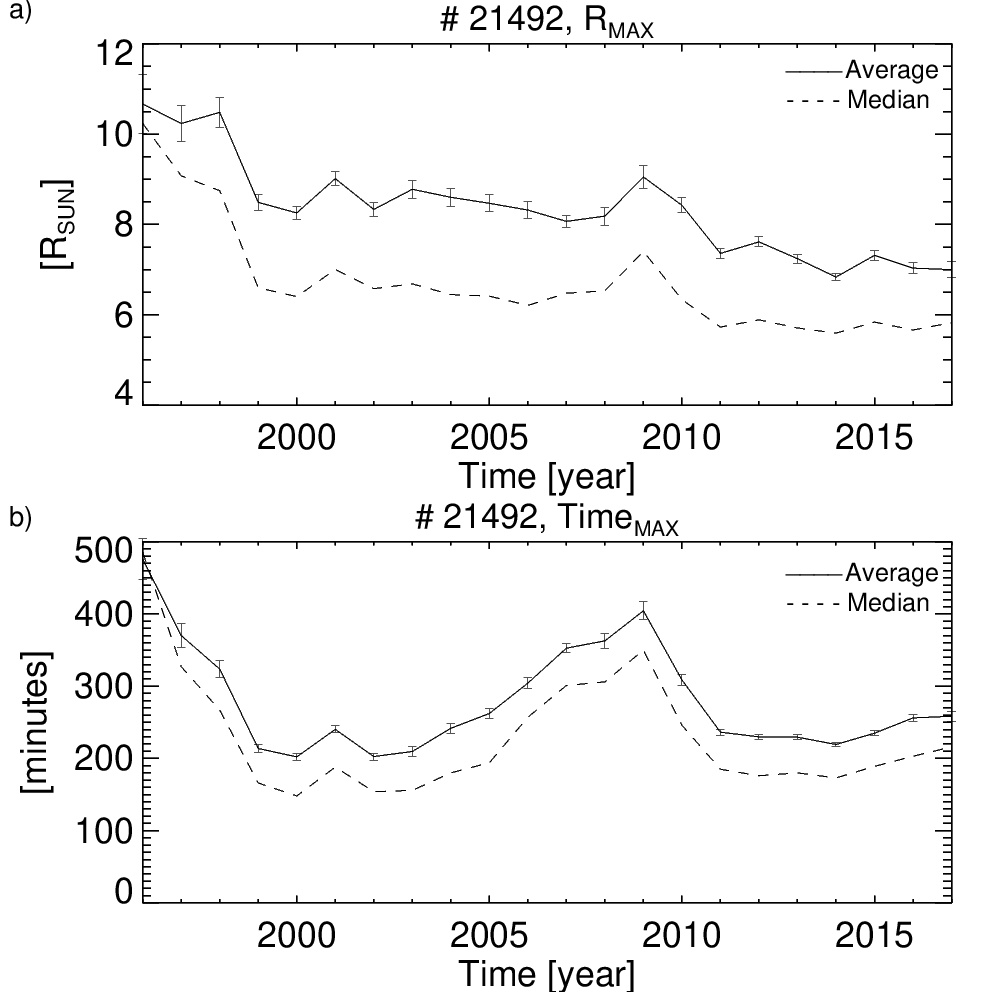}
  \caption{Average and median values of R$_{MAX}$ and Time$_{MAX}$ vs cycles of solar activity.}
             \label{FigGam}%
   \end{figure*}

\begin{figure*}
  \centering
  \includegraphics[width=16cm,height=10cm]{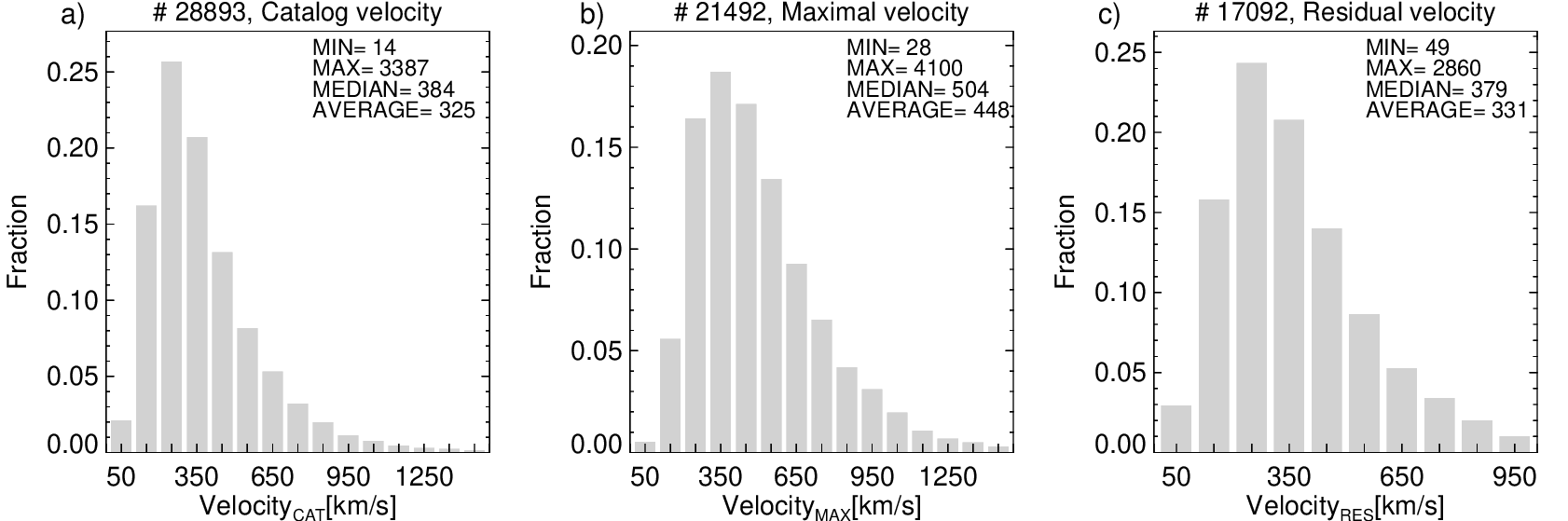}
  \caption{Distributions of the catalog velocity (Velocity$_{CAT}$, panel~a), maximum velocity (Velocity$_{MAX}$, panel~b), and residual velocity (Velocity$_{RES}$, panel~c) of the CMEs in the SOHO/LASCO catalog.}
            \label{FigGam}%
  \end{figure*}

Finally, it is important to analyze how these parameters (R$_{MAX}$, Time$_{MAX}$) change with activity of solar cycles. Figure~11 shows the variation in R$_{MAX}$ (panel~a) and Time$_{MAX}$ (panel~b) with solar cycles 23 and 24. In the case of R$_{MAX}$ we observe a systematic decline with time, from about 11 R$_{SUN}$ at the beginning of the 23rd solar cycle to about 7 R$_{SUN}$ at the end of the 24th solar cycle. This trend is disturbed by two small fluctuations in    the maximum of cycle 23 (1999 - 2000) and  in 2009. These results are intriguing as it suggests that the average propelling force (Lorentz force) driving CMEs  systematically decreases for more than two decades. The same trend is observed for the SSNs since cycle 22. The SSNs during the last solar maximum was about 50\% lower than during the 22nd cycle. Unfortunately, we do not have an appropriate observation of CMEs during the 22nd cycle. In the case of CMEs we can only compare data for the last two solar cycles.
As seen in figure 11, panel a, the average distance at which CMEs are accelerated is much smaller at the maximum of cycle 24 ($\approx$7 R$_{SUN}$) compared to cycle 23 ($\approx$9 R$_{SUN}$). This is another example showing different courses of solar activity in  various cycles. As previously stated, in cycle 24 the ejections are less energetic; therefore, their initial acceleration occurs at a smaller distance compared to cycle 23. Additionally, the R$_{MAX}$ is significantly high during the minimum of solar activity between cycles 22 and 23 (1996 - 1998). During this period, at the beginning of the catalog creation, several narrow events could have been missed (\citealt{Yashiro08}), which  means that during this period the catalog contains slow but clear and wide  ejections. According to results obtained by \citet{Sachdeva15, Sachdeva17}, such CMEs are characterized by the initial acceleration, which can last up to considerable distances from the Sun.

 Figure 11 panel b shows the Time$_{MAX}$ (the period in which the CMEs are significantly accelerated) following the trend of solar cycle activity. During the solar maximum, when the acceleration is dominated by the Lorentz force, Time$_{MAX}$ is shortest and on average is about 250 minutes. During the solar minimum, when they are slow and their acceleration is dominated by  the interaction with the solar wind, these times are approximately twice as long. Unlike R$_{MAX}$, the Time$_{MAX}$ behaves identically in solar cycle 23 and 24 despite the differences in the intensities (average Lorentz force) of both cycles.  In the 24th cycle, despite the lower average propelling force, CMEs expanded (in width) in the reduced total pressure (plasma + magnetic) of the interplanetary medium in which the drag force is much less than in the 23rd cycle.
 
 \begin{figure*}

\includegraphics[width=6.3cm,height=6cm]{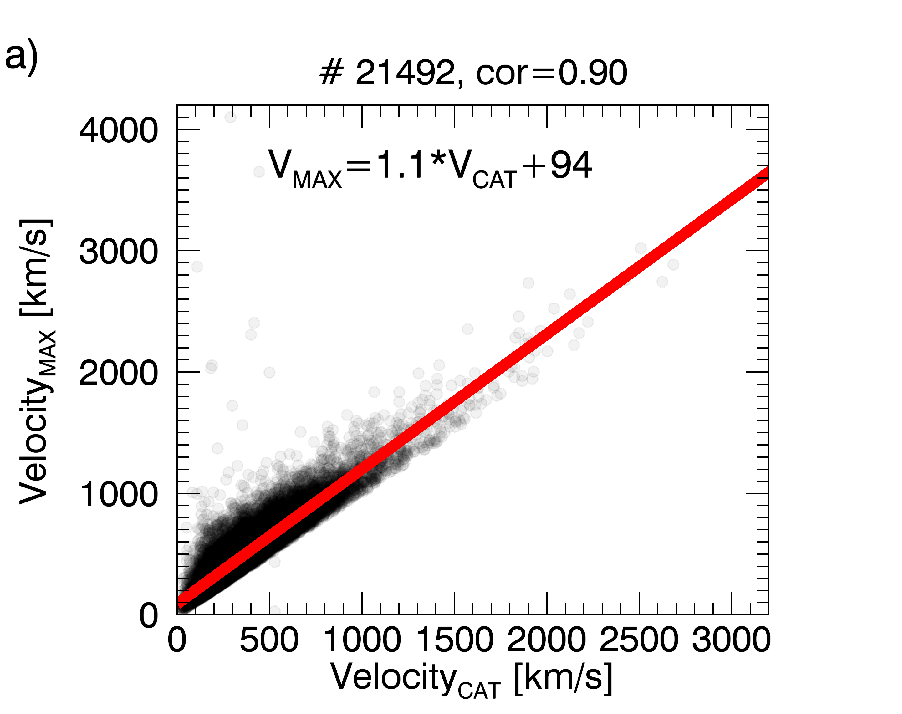}
\includegraphics[width=6.3cm,height=6cm]{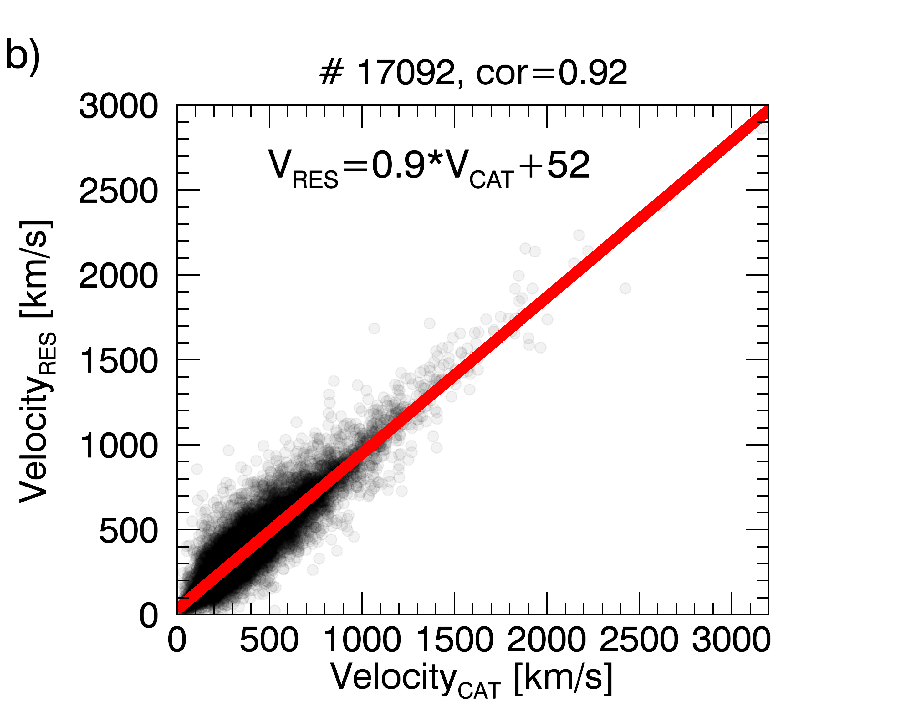}
\includegraphics[width=6.3cm,height=6cm]{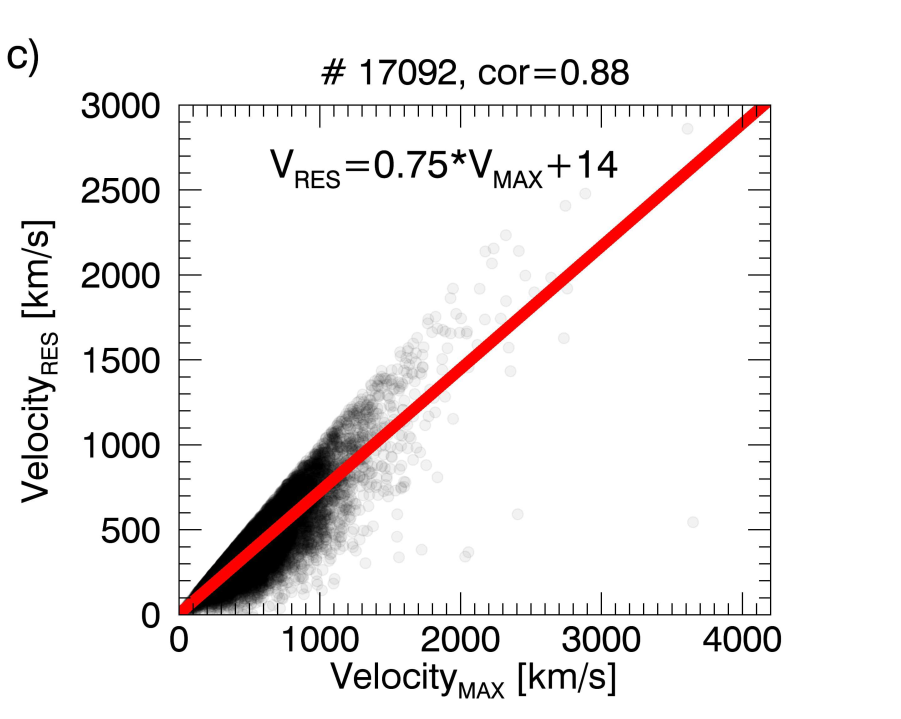}

\caption{Scatter plots of the catalog vs maximum velocity (panel a), the catalog vs residual velocity (panel b), and the maximum vs residual velocity (panel c) of the CMEs in the SOHO/LASCO catalog.}
\end{figure*}

\subsection{Velocities of CMEs}

In this section we describe the properties of different velocities characterizing propagation of CMEs. In Figure~12 we present distributions of the catalog (V$_{CAT}$, panel~a), maximum (V$_{MAX}$, panel~b), and residual (V$_{RES}$, panel~c) velocities (see Figure~1).  The catalog velocity is obtained by the linear fit to  all height-time measurements for a given CME; therefore, it reflects an average velocity of CME in the LFV. From the figure it is evident that the distributions of V$_{CAT}$  and V$_{RES}$ are similar. Median and average values for these parameters are about 380 and 330~km s$^{-1}$, respectively. They differ only in the range of the observed speeds. The maximum of the catalog and residual speeds are 3387 and 2860~km s$^{-1}$, respectively. At the same time, we  note that the maximum speeds differ significantly in panel b from those discussed above. Its median and average values are 504 and 448~km s$^{-1}$, respectively. The maximum speed from the figure is 4000~km s$^{-1}$ as our study is limited to CMEs with average speed $\approx$1500~km s$^{-1}$. However,  for fast events the maximum speeds may exceed 4000~km s$^{-1}$ indicating that CMEs can achieve significantly higher speeds than those estimated from the SOHO/LASCO catalog. It is important to note that the maximum instantaneous CME speed that we obtained is only a lower limit. There are  two reasons for this. First, LASCO coronagraphic observations are affected by the projection effect. Second, the fastest CMEs that we mentioned above are observed in coronagraphs only in two or three images. Hence, their instantaneous velocities cannot be determined and are not included in this study. We can assume that after taking into account these two effects (projection and selection) CMEs can reach maximum speeds of $\approx$6000~km~s$^{-1}$.

In Figure 13 the scatter plots show the relationship between V$_{CAT}$, V$_{MAX}$, and V$_{RES}$ for all the considered CMEs. We observe a significant correlation between these velocities (higher than 0.88). In the respective panels we  present equations describing the relations between these velocities.

At the end of this chapter we present the evolution of the catalog, maximum, and residual velocities along with cycles of solar activity. In Figure 14 we show changes of the yearly median and average  values of V$_{CAT}$ (panel~a), V$_{MAX}$ (panel~b), and V$_{RES}$ (panel~c) with solar cycles 23 and 24. Since these speeds are strictly correlated, their average and median values have a similar course over time, as seen in the figure. The average and median values of these velocities clearly follow the cycles of solar activity. As in the case of sunspots, we observe twin peaks during the solar maximum. However, this dual peak in the case of these velocities is different from that observed in the case of spots as the amplitudes of velocity peaks are opposite to the sunspots. If the first maximum is greater than the second in the case of sun spots, then for the velocities we observe the opposite situation. This was witnessed in cycle 23, but the situation is reversed for cycle 24. During the solar minimum, when the domination of drag force was seen in the LASCO  field of view, the average CME velocities determined by different methods were similar (about 300~km~s$^{-1}$). These differed during the solar maximum when the Lorentz force prevailed in the LFV. The maximum average values of these velocities are $\approx$550~km~s$^{-1}$, $\approx$600~km~s$^{-1}$, and $\approx$500~km~s$^{-1}$ for V$_{CAT}$, V$_{MAX}$, and V$_{RES}$, respectively.
As we can also see, the average and median annual values of the considered velocities are lower (by $\approx$100~km~s$^{-1}$) in the 24th cycle than in the  23rd cycle. This  reflects the decrease in the average Lorentz force during the 24th cycle.

\begin{figure*}
  \centering
  \includegraphics[width=16cm,height=12cm]{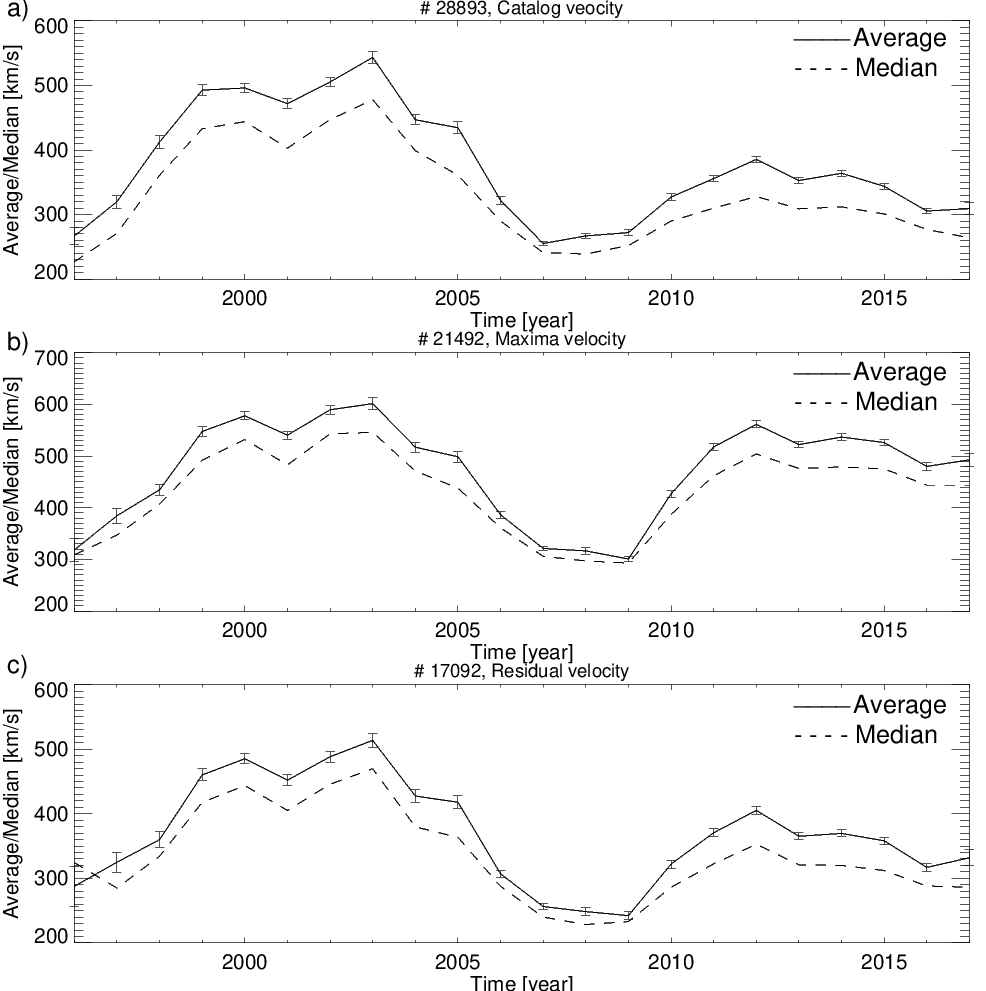}
  \caption{Yearly median and average values of the  catalog, maximum, and residual velocities vs time of the CMEs in SOHO/LASCO catalog. Error bars are the standard error on the average.}
             \label{FigGam}%
   \end{figure*}

\section{Discussion and conclusions}

In this study we evaluated the kinematics of CMEs during 23rd and 24th solar cycle. We have shown that the kinematics are more complex than  would appear from their characteristics included in the SOHO/LASCO catalog. In the LFV the expansion of CMEs are subject to two main phases: the initial and residual accelerations.
\begin{itemize}
  \item The initial acceleration is characterized by a rapid increase in the speed of CMEs up to V$_{MAX}$.  We demonstrated that the initial acceleration is in the range 0.24 - 2616~m s$^{-2}$ with median (average) value of 57~m s$^{-2}$ (34~m s$^{-2}$). It is evident that the initial acceleration can be quite significant, but our values are much lower compared to previous results of \citealt{Zhang06}.\\
  
  \item The residual acceleration is characterized by a gradual decrease in the speed of CMEs from V$_{MAX}$ to V$_{RES}$. The residual acceleration covers the range $-$1224 - 0~m s$^{-2}$ with median (average) value of $-$34~m s$^{-2}$ ($-$17~m s$^{-2}$). It is quite significant in comparison to previous studies of \citealt{Zhang06}.\\
  
  \item We found that the R$_{MAX}$ is in the range 2.5 - 28.1 R$_{SUN }$ with median (average) value of 7.8 R$_{SUN}$ (6.0 R$_{SUN }$) and the Time$_{MAX}$ is in the range 1.7 - 1092~minutes with median (average) value of 150 minutes (115) minutes. The obtained results allow us to conclude that CMEs are mostly accelerated up to the distance $\approx5.5$ R$_{SUN}$ within about two hours of the onset. This is in contradiction with previous studies that showed that the main acceleration phase takes place over a distance of about 3.3 R$_{SUN}$ (\citealt{Zhang04}).\\

      \item The analysis shows that the R$_{MAX}$ depends on the velocity of CMEs (Figure~9 and 10, panel~c). Only slow events can be accelerated close to the Sun, within the distance $\geq$2.5 R$_{SUN}$  from the Sun. The median and average values of R$_{MAX}$ for slow events are 6.3 and 5.0 R$_{SUN}$, respectively.  The faster events need greater acceleration distance to achieve the maximum speed. The minimum acceleration distance for these CMEs is about 3.5 R$_{SUN}$. The average and median values of R$_{MAX}$ for these events  are 11.0 and 9.1 R$_{SUN}$, respectively. The findings from this paper differ from those of  the previous study by \citet{Sachdeva15, Sachdeva17} as a greater number of events were considered in our analysis.\\

          \item It should also be clearly noted that the significant driving force (Lorentz force) can operate up to the distance 6~R$_{SUN}$ from the Sun and during the first two hours of propagation (Figure~10).\\
          
    \item  Our study revealed that the considered parameter (ACC$_{INI}$, ACC$_{RES}$, Time$_{MAX}$, and V$_{MAX}$) mostly follow the cycles of solar activity and the intensities of individual cycles.\\

        \item Our analysis has shown a sudden significant drop in the average residual acceleration in the maximum of cycle 24. It seems
that this is due to the anomalous behavior of CMEs in the 24th solar
cycle.  During this cycle, CMEs are less
energetic but they still expand (in width) rapidly in the reduced total pressure (plasma + magnetic) of the interplanetary
medium. This is due to the rapid decrease in the propelling force (Lorentz
force) as the internal magnetic field dilutes quickly and the drag force significantly prevails over the
propelling force compared to the maximum of solar cycle 23 (Figure~5, panel~c). \\

        \item In the case of R$_{MAX}$, we observe systematic decline with time, from about 11 R$_{SUN}$ at the beginning of the 23rd solar cycle to about 7 R$_{SUN}$ at the end of the 24th solar cycle.  This suggests that the average propelling force (the average Lorentz force) driving CMEs, for more than two decades, systematically decreased. The same trend is observed for the SSNs since cycle 22. The SSNs during the last solar maximum was about 50\% lower than during the 22nd cycle. Unfortunately, we do not have appropriate observations of CMEs during the 22nd cycle. In the case of CMEs we can only compare data for the 23rd and 24th solar cycles (Figure~11).
\end{itemize}

The presented  statistical analysis reveals that at the beginning of their expansion, in the vicinity of the Sun, CMEs are subject to several factors (Lorentz Force, CME-CME interaction, speed differences between leading and trailing parts of the CME) that determine their propagation. Although their average values of catalog accelerations are always close to zero, a more detailed study shows that their instantaneous accelerations may be quite different depending on the conditions prevailing in the Sun and the environment in which they propagate. These conditions vary depending on the individual eruption and over time as the solar activity changes.

\begin{acknowledgements}
    Anitha Ravishankar and Grzegorz Micha$\l$ek were supported by NCN through the grant UMO-2017/25/B/ST9/00536 and DSC grant N17/MNS/000038. This work was also supported by NASA LWS project led by Dr. N. Gopalswamy.\\
    The authors thank the referee for the useful comments and suggestions that have greatly improved the quality of the manuscript. We thank all the members of the SOHO/LASCO consortium who built the instruments and provided the data used in this study. LASCO images are courtesy of SOHO consortium. This CME catalog is generated and maintained at the CDAW Data Center by NASA and The Catholic University of America in cooperation with the Naval Research Laboratory. SOHO is a project of international cooperation between ESA and NASA. 
\end{acknowledgements}

%
%

\end{document}